\documentclass[%
 reprint,
 amsmath,amssymb,
 aps,
]{revtex4-2}

\usepackage{graphicx}
\usepackage{dcolumn}
\usepackage{bm}

\begin{document}

\preprint{APS/123-QED}

\title{Feasibility of Liquid-phase Xenon Proportional Scintillation for Low-energy Physics}

\author{Jianyang Qi}
 \email{Corresponding author: jiq019@ucsd.edu}
 \affiliation{University of California, San Diego}
\author{Kaixuan Ni}%
 \email{Corresponding author: nikx@physics.ucsd.edu}
\affiliation{University of California, San Diego}%
\author{Haiwen Xu}%
\affiliation{University of California, San Diego}%
\author{Yue Ma}%
\affiliation{University of California, San Diego}%
\author{Yuechen Liu}%
\affiliation{University of California, San Diego}%
\date{\today}

\begin{abstract}
Dual phase xenon time projection chambers (TPCs) detect both the scintillation photons and ionization electrons created by energy depositions within the liquid xenon (LXe) volume. The electrons are extracted from the interaction site through a gas gap, where they meet a high electric field where proportional scintillation occurs. This converts the electron signal into a light signal, and yields a high electron detection efficiency with a gain of tens of photoelectrons (PE) per electron. This technique of detecting both scintillation and ionization gives dual phase xenon TPCs the capability to distinguish between electronic and nuclear recoils, which is a key part of how these detectors are able to reach world-leading limits on Weakly Interacting Massive Particle (WIMP) dark matter. However, not all electrons can be extracted through the liquid-gas interface, and a constant millimeter-scale gas gap needs to be maintained, which may be a technological challenge if dual-phase xenon TPCs are to be scaled up for future dark matter searches. Furthermore, there is a background of single-electron peaks that follow a large ionization signal (S2) of unclear origin which may be due in part to the liquid-gas interface, and limits the sensitivity of these detectors towards low mass dark matter. In this paper, we demonstrate that a purely single-phase liquid xenon TPC which produces proportional scintillation directly in the liquid is still capable of discriminating between electronic and nuclear recoils, but that the background of single-electrons following an S2 is still likely unrelated to the liquid-gas interface.
\end{abstract}

\maketitle


\section{\label{sec:introduction}Introduction}
Dual-phase xenon time projection chambers (Xe TPCs) are widely used and optimized to detect dark matter, primarily in the form of Weakly Interacting Massive Particles (WIMPs). These WIMPs are expected to create low-energy (keV$_{ee}$ scale) recoils on large xenon nuclei, and their detection requires a detector to have both a low energy threshold as well as an ability to discriminate nuclear recoils (NR) signals from electronic recoil (ER) backgrounds. Dual-phase Xe TPCs can overcome both of these challenges. The required low energy threshold can be reached \cite{LUX:2017ojt,Lenardo:2019fcn} by detecting single-electrons using gas-phase proportional scintillation to amplify a single-electron to tens of photoelectrons (PE) \cite{XENON:2022ltv,LZ:2022lsv,PandaX-4T:2021bab}. Meanwhile, their ability to produce both prompt scintillation (S1) and ionization (S2) signals allows one to use the S2/S1 ratio to discriminate between ERs and NRs. With this, the current lowest upper-limits on the spin-independent WIMP-nucleon cross section are all placed by dual-phase xenon time projection chambers (TPCs) \cite{PandaX-4T:2021bab,LZ:2022lsv,XENON:2023cxc}. In addition to WIMP dark matter, if dual-phase Xe TPCs drop the S1 requirement and search for physics using only the ionization signal (i.e. ``S2-only"), their energy threshold is lowered and they may also be sensitive to lower mass dark matter such as sub-GeV WIMPs \cite{XENON:2019gfn,PandaX:2022xqx}, as well as Coherent Elastic Neutrino Nucleus Scattering (CE$\nu$NS) from reactor neutrinos \cite{Ni:2021mwa}.

However, the current generation of the largest dual phase Xe TPCs (LZ, PandaX-4T, and XENONnT) are unable to extract 100\% of the electrons from the liquid into the gas \cite{XENON:2022ltv,LZ:2022lsv,PandaX-4T:2021bab}. As a result, an energy deposition which results in only a few electrons may suffer considerable signal loss and resolution, thus hindering the ability to search for physics in S2-only channel. In addition, the next generation of dual-phase Xe TPCs will require even wider electrodes than the ones in LZ, PandaX-4T, and XENONnT at the time of this article. This entails the technological challenge of maintaining a few mm level gas gap over a distance of a few meters. Furthermore, these dual phase xenon TPCs have a known background of single-electrons that follow a large ionization signal (S2) both temporally and spatially, that can last up to one second after the S2 \cite{XENON:2021qze,LUX:2020vbj}. This background is referred to as an ``electron train", and is a particularly important limitation for the aforementioned S2-only searches.

Instead of amplifying electrons using gas xenon (GXe) proportional scintillation, one may instead amplify electrons using liquid xenon (LXe) proportional scintillation \cite{Masuda:1978tjp,Wang:1998gq,Aprile:2014ila,Lin:2021izy,Wei:2021nuk,Qi:2023bof,Martinez-Lema:2023zjk}. In doing so, the liquid-gas interface is eliminated -- along with the need to maintain a gas gap over a potentially large area -- and the extraction efficiency is effectively 100\%. However, this is typically at the cost of a low single-electron gain of $\mathcal{O}$(1) PE/electron, as the threshold electric field needed to create proportional scintillation in LXe is approximately 412~kV/cm \cite{Aprile:2014ila}, compared to the 1.3~kV/cm/bar \cite{Monteiro:2007vz} in GXe. In this paper, we show experimental evidence that a LXe TPC which produces proportional scintillation directly in LXe is still capable of discriminating between ER and NR despite its low single-electron gain, thus showing the potential feasibility of LXe proportional scintillation in WIMP dark matter searches. Furthermore, we provide insight on the origin of the electron trains background by showing that this background is still present even without a liquid-gas interface, thus showing its limitations for S2-only searches.





\section{\label{sec:level2}Detector Setup and Data Acquisition}

The detector used in this paper, which we call the Liquid Xenon Proportional Scintillation Counter (LXePSC), is a single-phase liquid xenon detector with a cylindrical inner volume that produces proportional scintillation via a thin, 17.8~$\mu$m diameter anode wire along its central axis, and twenty evenly spaced cathode wires parallel to the anode placed 2.5~cm away at the inner edge of the cylindrical barrel. The light from this detector is then seen by eight Hamamatsu R8520 PMTs that surround the inner barrel of the LXePSC, four on the top row and four on the bottom row . This detector geometry was first proposed by \cite{Lin:2021izy}, and the details of our detector hardware is described in our previous two papers \cite{Wei:2021nuk,Qi:2023bof}. In this run, we changed the anode from a 10~$\mu$m diameter wire to an 17.8~$\mu$m diameter wire in an attempt to achieve a higher single-electron gain. This is because if two anode wires have the same electric field at their surface, then a thicker wire will have a larger proportional scintillation region, as the thickness of the proportional scintillation region, $r_{PS}$, is given by \begin{equation}
    r_{PS} = (E_a/E_T - 1)r_a.
    \label{eq:r_prop_scint}
\end{equation} Here $E_a$ is the electric field at the anode, $E_T$ is the threshold electric field to start producing proportional scintillation, and $r_a$ is the radius of the anode.

\begin{figure}
    \centering
    \includegraphics[width = 0.99\columnwidth]{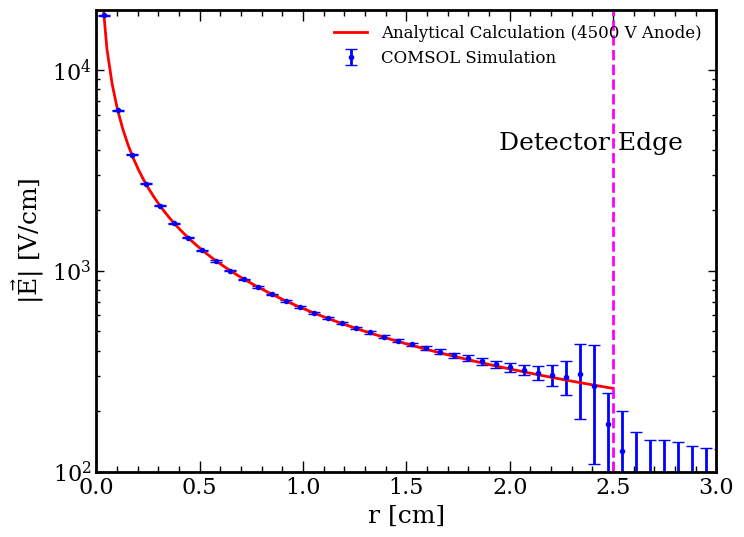}
    \caption{A comparison between the electric field simulation from COMSOL{\texttrademark} multiphysics and the analytical electric field $\Delta V/\ln(r_c/r_a)r$. Here, $\Delta V = V_a-V_c$ where $V_a=4.5$~kV and $V_c = -0.6543$~kV, $r_a = 17.8$~$\mu$m, and $r_c=2.5$~cm. The electric field simulation is sliced in the central 2~cm in $z$ of the detector, and the error bars are the standard deviation of the electric field due to inhomogeneities in $E_\theta$ and $E_z$ at a given $r$.}
    \label{fig:e_vs_r}
\end{figure}

In addition to changing the anode diameter, we also upgraded our data acquisition system to be able to take data in three different modes, with each mode optimized for each type of data. Background data and activated xenon data was taken using a CAEN V1725S digitizer running with the Dynamic Acquisition Window (DAW) firmware. This mode of data taking acquires \textit{every} pulse for \textit{every} channel which lies above a certain threshold, and is also called ``triggerless" or ``continuous" data taking. Data taken with DAW is best suited for low event rates, as high event rates can cause missing data in some PMT channels, but not others. The NR and ER data were taken with high-activity calibration sources in or near the detector which required us to use globally triggered data taking to use PMT coincidences to trigger on physical events. As such, the NR data was taken using the default waveform recording firmware. Afterwards, we upgraded our DAQ software to take data using Zero-Length Encoding (ZLE), which was used for the tritium ER data. This ZLE feature still records event windows with a global trigger, but suppresses the baseline by only saving pulses above $\sim$0.8 PE. Throughout this paper, \textit{pulses} refer to per-channel signals, while \textit{peaks} refers to pulses that are merged across channels. S1s and S2s are classified based on the time it takes for the peak to rise from 10\% to 90\% of its maximum height, the rise time, where S1s have a rise time of less than 25~ns, and S2s have a rise time greater than 25~ns and an area greater than 10~PE. We pair S1s and S2s into \textit{events} by starting from the last S2 in a dataset, and looking back 25~$\mu$s from its start time to find corresponding S1s. The largest S1 and S2 in this time window are called the main S1 and S2, and the second largest are called the alternate S1 and S2.

\subsection{\label{subsec:calibration_summary}Calibrations}
During this run, we took calibration data using two radioactive sources, a $^{252}$Cf spontaneous fission source and a tritiated methane low-energy beta source. NR data was taken with the $^{252}$Cf source placed behind approximately 40~cm of lead outside of the outer insulating vacuum jacket of the detector vessel. This is to improve the ratio of NR events to the high energy gamma rays which are also produced by the $^{252}$Cf source, as the attenuation length of $\mathcal{O}$(MeV) gammas in lead is less than the scattering length of $\mathcal{O}$(MeV) neutrons \cite{Brown20181,204146}. Afterwards, we moved the $^{252}$Cf source to a position close to the detector to activate the LXe and get two meta-stable neutron activated xenon isotopes, $^{129\rm m}$Xe and $^{131\rm m}$Xe, which have a half life of 8.8 and 11.8~days and produce 236 and 164~keV gammas, respectively \cite{Ni:2007ih}. These gamma lines were then used to measure the detector light detection efficiency ($g_1$) and single-electron gain ($g_2$), as described in Section \ref{sec:xeact}. In addition, tritiated methane was injected into our detector via our gas system, and was used to obtain ER data from tritium beta decays. The ER data and NR data are used together to estimate the leakage of ER events below the NR band median in $(S1, \log_{10}(S2/S1))$ space as described in Section \ref{sec:ernr}. In addition to calibration data, we took background data before any sources were introduced, which gave us, on average, 16~ms between consecutive events and allowed us to characterize the electron trains background in our detector, described in Section \ref{sec:etrains}.

\section{\label{sec:xeact}Activated Xenon Lines}
Two long lived metastable isotopes $^{129\rm m}$Xe and $^{131\rm m}$Xe were produced in our LXe volume by placing a $^{252}$Cf source with an activity of $\sim1.8\times10^6$~neutrons/s approximately 4~inches away from the center of the detector for two days. After we took the source away, the two metastable isotopes remained and gave us 236 and 164~keV gamma lines, respectively, for which we took triggerless data for a week. During this run, the anode was kept at a voltage of 4.5~kV and the cathode was kept at a voltage of -0.6543~kV. The cathode voltage was chosen as the average of the PMT voltages while the anode voltage is the highest achievable voltage before the onset of spurious light emission (see Subsection \ref{subsec:light_emission}). A comparison of the simulated electric field using COMSOL\textsuperscript{\texttrademark} multiphysics and the analytical calculation for the central $\sim$2~cm in $z$ of the detector is shown in Figure \ref{fig:e_vs_r}.



\begin{figure*}
    \centering
    \includegraphics[width = 0.329\textwidth]{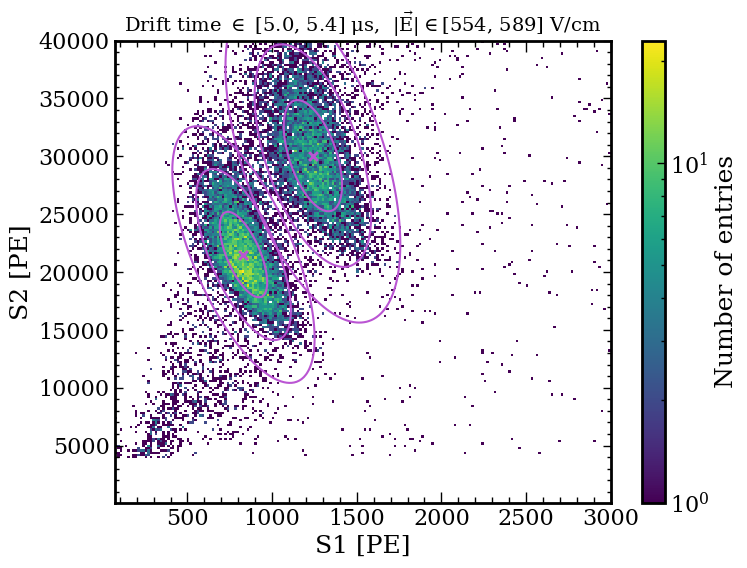}
    \includegraphics[width = 0.329\textwidth]{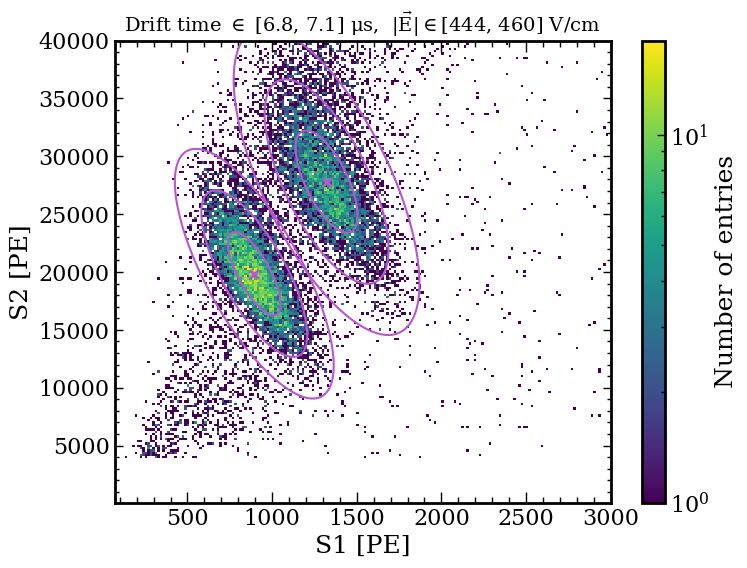}
    \includegraphics[width = 0.329\textwidth]{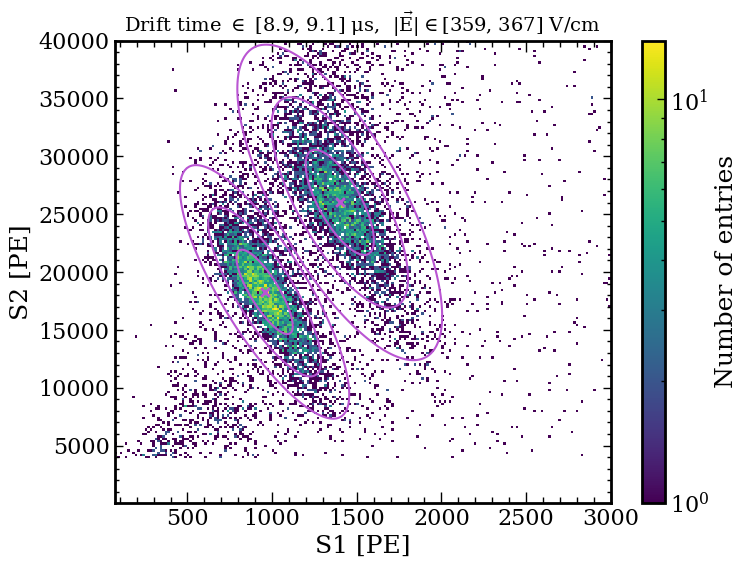}
    \caption{Distributions of the $^{129\rm m}$Xe and $^{131\rm m}$Xe lines in $(S1,S2)$ space after the cuts and corrections described in Subsection \ref{subsec:cuts_corrections_xeact}. The 164~keV $^{131\rm m}$Xe line is the bottom distribution while the 236~keV $^{129\rm m}$Xe line is the top distribution. The center S1 and S2 ($S1_c$ and $S2_c$) are shown with an ``x" marker and is fit with a 2-d Gaussian.}
    \label{fig:xenon_lines_dt_slice}
\end{figure*}

The LXePSC's radial, $\propto \hat{r}/r$ electric field means that the radial coordinate of an event, $r$, can be inferred from the time between the S1 and the S2, known as the drift time. By slicing the events into drift time bins, we can obtain a measurement of the mean S1 and S2 of the $^{129\rm m}$Xe and $^{131\rm m}$Xe lines for each $r$ slice, which in turn corresponds to slices in the electric field magnitude. Figure \ref{fig:xenon_lines_dt_slice} shows the distributions of the $^{129\rm m}$Xe and $^{131\rm m}$Xe lines in $(S1,S2)$ space for a few drift time slices. As the electrode conditions are kept fixed, the average light collection efficiency ($g_1$) and single-electron gain ($g_2$) are also fixed despite the changing electric field in the LXe bulk. For a fixed energy, the S2 (S1) size increases (decreases) with the electric field due to the decrease in the probability for electrons from an ionization to recombine with a Xe atom \cite{Doke:2002oab}. However, the total number of quanta (electrons plus photons) is independent of the electric field. As such, for a monoenergetic energy deposition, the corrected mean S1 and S2 in each drift time slice, $cS1_c$ and $cS2_c$, should lie on the line $E = W(cS1_c/g_1+cS2_c/g_2)$. Here, the corrections are to correct for position related inhomogeneities not related to the electric field. This energy is commonly referred to as the ``Combined Energy Scale" (CES). Fitting a line through the scatter plot of $cS1_c/E$ and $cS2_c/E$ (i.e. the light and charge yield) gives the parameters $g_1$ and $g_2$ for a given $W$ value, and is commonly referred to as a Doke plot \cite{Doke:2002oab}.

\subsection{\label{subsec:cuts_corrections_xeact}Systematic Uncertainties, Corrections, and Cuts}

In the beginning of this run, we performed a PMT calibration, using a method similar to our previous run \cite{Qi:2023bof}, but found that the new PMT gains show a seemingly unphysical z-distribution that is not centered around 0. The z-distribution is inferred from the S2-asymmetry which is given by \begin{equation}
    S2_{asym} = \frac{S2_{\text{top 4 PMTs}}-S2_{\text{bottom 4 PMTs}}}{S2_{\text{top 4 PMTs}}+S2_{\text{bottom 4 PMTs}}}
\end{equation} as described in Ref. \cite{Wei:2021nuk}. This shift in the S2-asymmetry ($z$) distribution, leads us to use our old set of PMT gains from the previous run for the remainder of this paper, aside from $g_1$ and $g_2$, where we still include the choice of the PMT gain set as a systematic uncertainty. None of the other results in this run are dependent on the absolute pulse areas (i.e. the PMT gains).

S2 peaks can sometimes saturate the 2~V dynamic range of the digitizer which can lead to underestimated S2 areas. However, oftentimes, not all of the channels are saturated, in which case we can use the sum of the channel waveforms for the unsaturated channels, to estimate what the waveform of the saturated channels would be, had they not been saturated. The saturation correction procedure is detailed in Appendix \ref{appendix:satcor}.

In addition, the S1 needs to be, in principle, corrected for the $r$-dependent light collection efficiency (LCE), and S2 needs to be corrected for the charge loss due to electrons attaching to electronegative impurities while drifting to the anode. The S1 LCE can be obtained from an optical simulation using GEANT4. From this, we can multiply the S1 in each drift time slice by the ratio of the mean S1 LCE and the S1 LCE for the particular drift time, $\langle LCE_{S1}\rangle/LCE_{S1}(t_d)$. In typical TPCs which have nearly uniform drift field, the S2s are usually corrected by $\exp(t_d/\tau_e)$ where $t_d$ is the drift time and $\tau_e$ is the electron lifetime \cite{XENON:2019gfn}. However, in the cylindrical LXePSC, the raw S2 area also varies at different radii, as the recombination of the excitons depends of drift field. This effect will be referred to as the ``recombination-drift time coupling" effect. Likewise, the recombination-drift time coupling effect will also affect S1s, which makes it difficult to obtain a validation for the S1 correction. As such, we include a systematic uncertainty for whether or not the S1 LCE correction is used. S2 area can be modelled as \begin{equation}
    S2 = n_e(|\textbf{E}|, \epsilon)e^{-dt/\tau_e}g_2
\end{equation} where $|\textbf{E}|$ is the magnitude of the electric field and $\epsilon$ is the deposited energy. To estimate the electron lifetime, $n_e$ can be estimated using NEST given the local electric field inferred from the drift time, and the energy of either the $\epsilon$=164~keV or 236~keV line. The electron lifetime is fit for each day of the activated xenon data taking, and is estimated to be about 80-130~$\mu$s. The ambiguity in the electron lifetime leads us to include it as a systematic uncertainty, rather than a correction, whereby we compute the systematic uncertainty in $g_1$ and $g_2$ by varying $\tau_e$ between 80-130~$\mu$s.

\begin{figure}
    \centering
    \includegraphics[width=0.99\columnwidth]{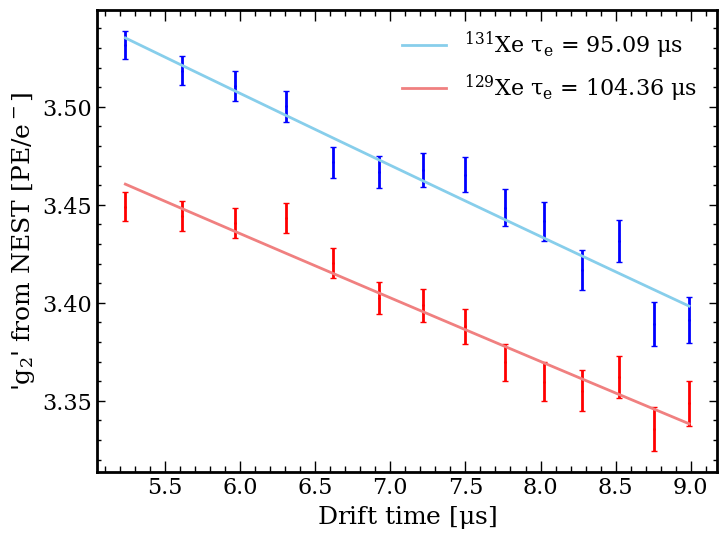}
    \caption{$S2_c/n_e$ as a function of drift time. This is referred to as '$g_2$' (with quotations) since it's similar to $g_2$ but with the electron lifetime factor $e^{-dt/\tau_e}$. We can see that $\tau_e$ from $^{131\rm m}$Xe differs from $^{129\rm m}$Xe.}
    \label{fig:nest_elife}
\end{figure}

To select for clean activated xenon events, we select for single-scatter events in a near the center of the detector. Here, events in the center of the detector have an S2 asymmetry between -0.25 and 0.25, which corresponds to the central 2~cm of the detector in $z$, far from wall backgrounds and fringe electric fields.


\subsection{\label{sec:g1g2}$g_1$, $g_2$, and Doke Plot}

\begin{figure}
    \centering
    \includegraphics[width=0.99\columnwidth]{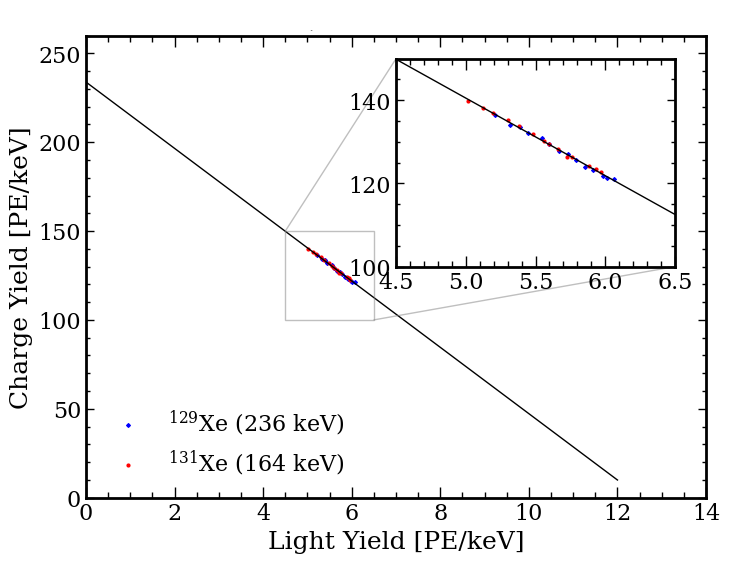}
    \caption{Doke plot for an assumed electron lifetime of 100~$\mu$s. This data uses the old PMT gains and the S1 LCE correction. We can see clearly that both the $^{129\rm m}$Xe and $^{131\rm m}$Xe light and charge yields lie on the same line.}
    \label{fig:doke_plot}
\end{figure}

\begin{figure*}
    \centering
    \includegraphics[width=0.99\textwidth]{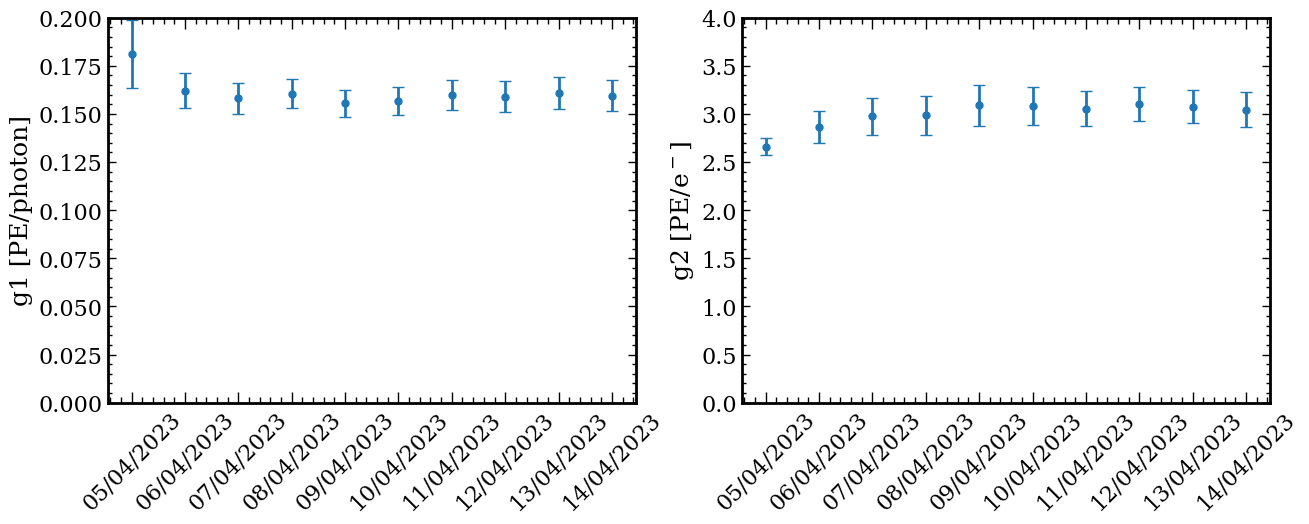}
    \caption{$g_1$ and $g_2$ obtained from the activated xenon lines plotted for each day of data taking. The data from April 5th seems to be an outlier and is not indicative of any trend, but we include it for completeness.}
    \label{fig:g1_g2_plot}
\end{figure*}

An example Doke plot from data taken on April 12th, 2023 is shown in Figure \ref{fig:doke_plot}. Here, the cuts and corrections to $S1_c$ and $S2_c$ (Figure \ref{fig:xenon_lines_dt_slice}) from the previous section are applied. The charge and light yields are $QY=cS2_c/E$ and $LY=cS1_c/E$, and are plotted for both activated xenon lines. Different points correspond to different drift time slices, and we see that the points are indeed anti-correlated. The slope and y-intercept of the Doke plot are $-g_2/g_1$ and $g_2/W$, respectively. $g_1$ and $g_2$ are obtained with $W=13.5$~eV \cite{szydagis_2018_1314669} and are shown with all systematic uncertainties included in Figure \ref{fig:g1_g2_plot}.

By averaging over the last six days of the activated xenon calibration when the $g_1$ and $g_2$ plots seem to stabilize, we estimate $g_1=0.159\pm0.008$~PE/photon and $g_2=3.07\pm0.19$~PE/electron. These estimates and their corresponding systematic uncertainties include the ambiguity from both sets of PMT gains. However, for the reasons stated at the beginning of this section, the $g_1$ and $g_2$ used to do simulations and corrections, which are obtained using the old set of PMT gains, are instead $g_1=0.162\pm0.007$~PE/photon and $g_2=3.2\pm0.1$~PE/electron.

\section{\label{sec:ernr}ER/NR Discrimination}

\begin{figure*}
    \centering
    \includegraphics[width=1\textwidth]{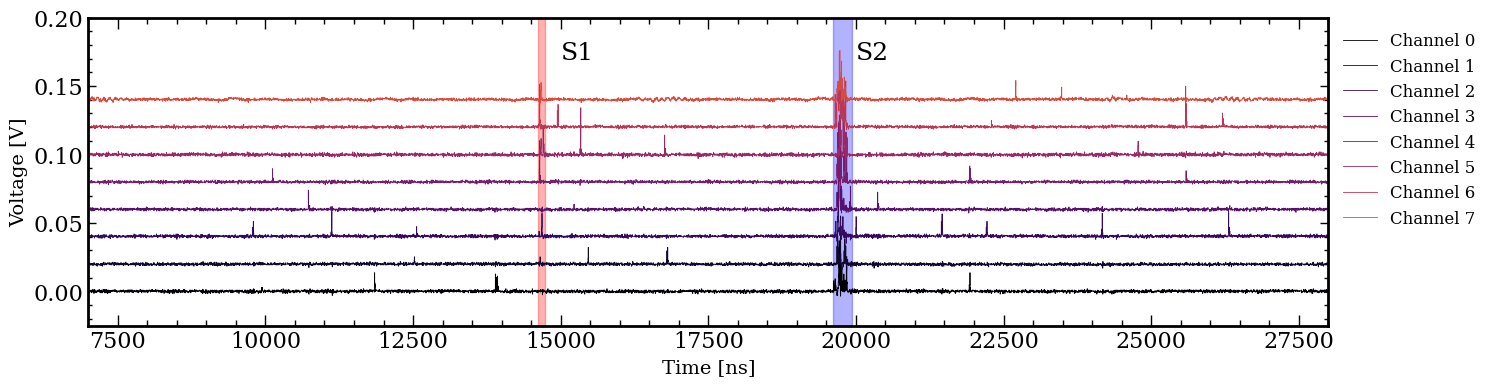}
    \includegraphics[width=1\textwidth]{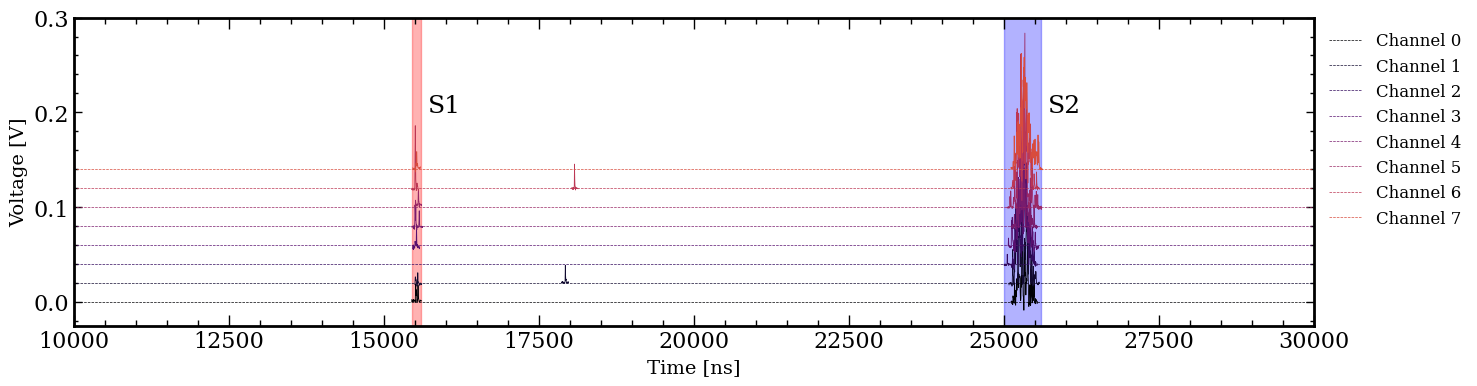}
    \caption{An example NR (top) and tritium (bottom) waveform after the cuts described in Appendix \ref{appendix:er_nr_corrections}. The NR data was taken with the waveform acquisition firmware and thus shows a baseline, while the tritium data was taken with ZLE and thus the baseline is indicated with dashed lines.}
    \label{fig:er_nr_wfs}
\end{figure*}
\begin{figure*}
    \centering
    \includegraphics[width=0.49\textwidth]{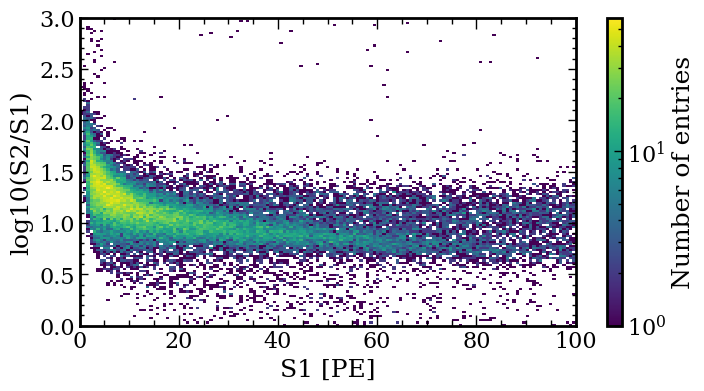}
    \includegraphics[width=0.49\textwidth]{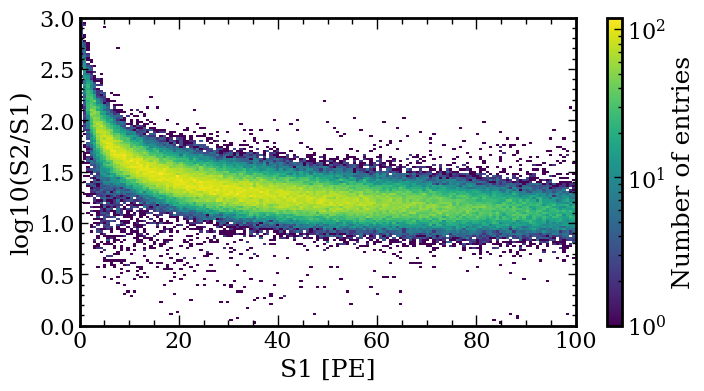}
    \includegraphics[width=0.49\textwidth]{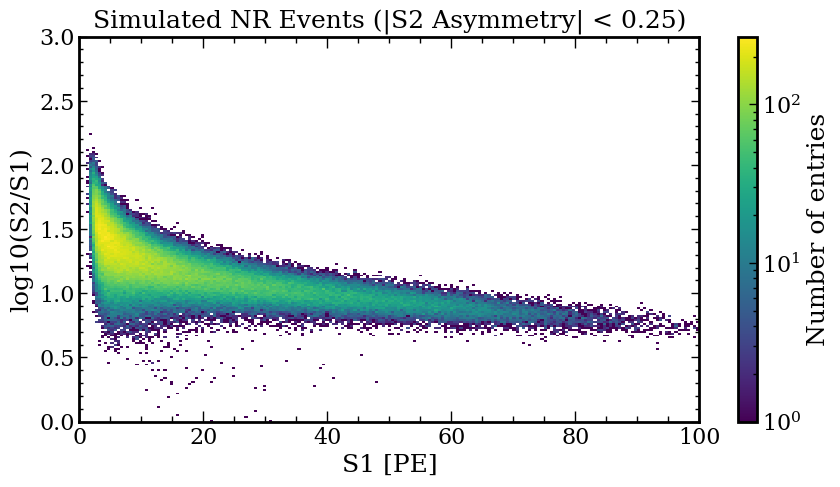}
    \includegraphics[width=0.49\textwidth]{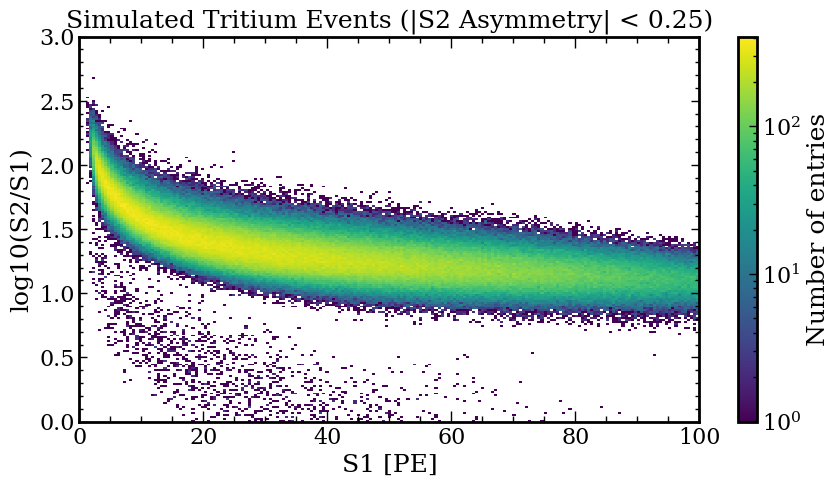}
    \caption{Top row: The $^{252}$Cf NR and tritium ER bands from \textit{data} after all cuts and corrections are applied. The NR data was taken along with multiple higher energy gammas, which appear at higher values of S1 and $\log_{10}(S2/S1)$. Bottom row: The simulated $^{252}$Cf NR and tritium ER bands after applying the S2 asymmetry cut. These bands were simulated using the method described in Section \ref{subsubsec:wfsim}, and also features leakage events largely due to the charge insensitive volume.}
    \label{fig:er_nr_bands}
\end{figure*}

The NR data was taken from March 23rd, 2023 to March 30th, 2023 while the tritium ER data was taken from June 1st, 2023 to June 7th, 2023. In between this time, we took activated xenon data, after which we experienced a vacuum failure on our outer vacuum jacket. This forced briefly stop operations, including continuous purification of the Xe. When we turned our purification back on, we switched the circulation speed from 3~standard-liter-per-minute (SLPM) to 5~SLPM, which actually significantly increased our S2 areas. As such, the tritium data and NR data are not directly comparable, as the electron lifetime and $g_2$ are not the same. To correct for this, we normalize our S1 and S2 areas for the ER and NR data to what their values would be if the data was taken on April 12th, 2023 -- one of our activated xenon days. We describe our methodology to correct and select for clean NR and tritium events in Appendix \ref{appendix:er_nr_corrections}. An example waveform of a tritium ER and an NR event after all of these cuts is shown in Figure \ref{fig:er_nr_wfs}, and the ER and NR bands after all cuts and corrections are shown in Figure \ref{fig:er_nr_bands}.

\subsection{\label{subsec:leakage}Leakage Estimations}

\begin{figure*}
    \centering
    \includegraphics[width=0.32\textwidth]{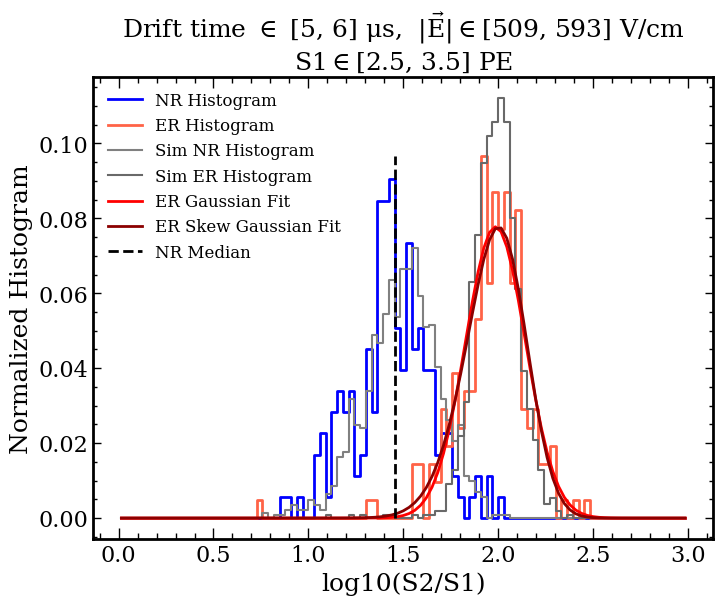}
    \includegraphics[width=0.32\textwidth]{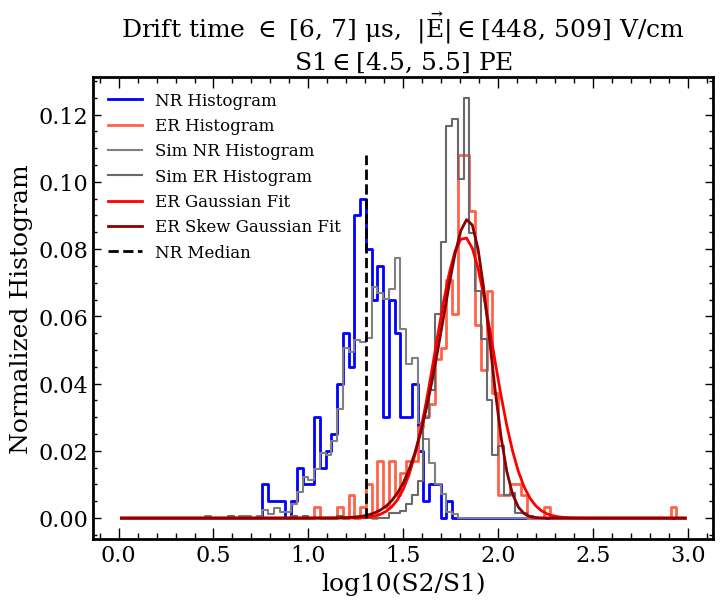}
    \includegraphics[width=0.32\textwidth]{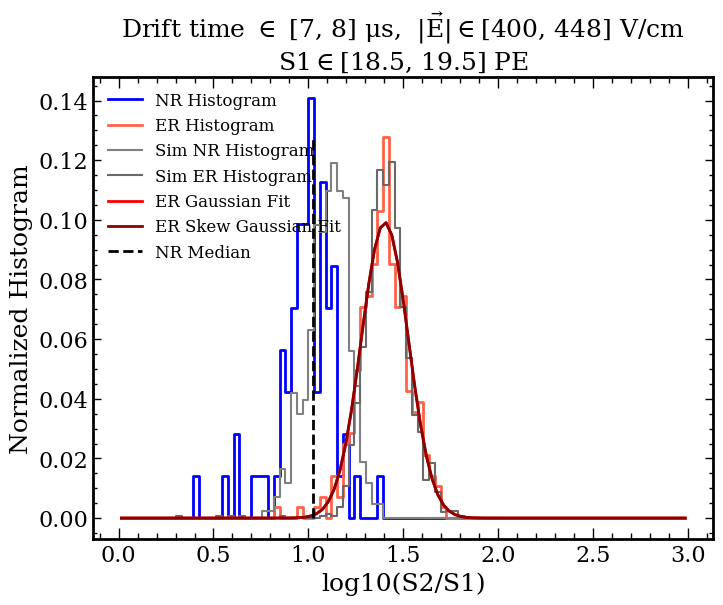}
    \caption{Example fits used to calculate the leakage per S1 bin for a variety of drift time and S1 slices as indicated by the plot titles. The fits are to the section of the tritium histogram with at least 3 counts in each bin. Simulated ER and NR distributions that are sliced in drift time in $\log_{10}(S2/S1)$ are shown in gray. While the tritium distribution matches nicely for larger S2s, the simulation shows some mismatch towards lower S2s. Furthermore, the NR distribution matches nicely for small S1s, but is mismatched for larger S1s. This mismatch can arise from a multitude of reasons such as the inherent systematic uncertainty of NEST, the mis-modelling of the optical simulation, or even differences between the simulated drift time and true drift time in their correspondence to $r$.}
    \label{fig:example_fits_leakage}
\end{figure*}

\begin{figure}
    \centering
    \includegraphics[width=0.98\columnwidth]{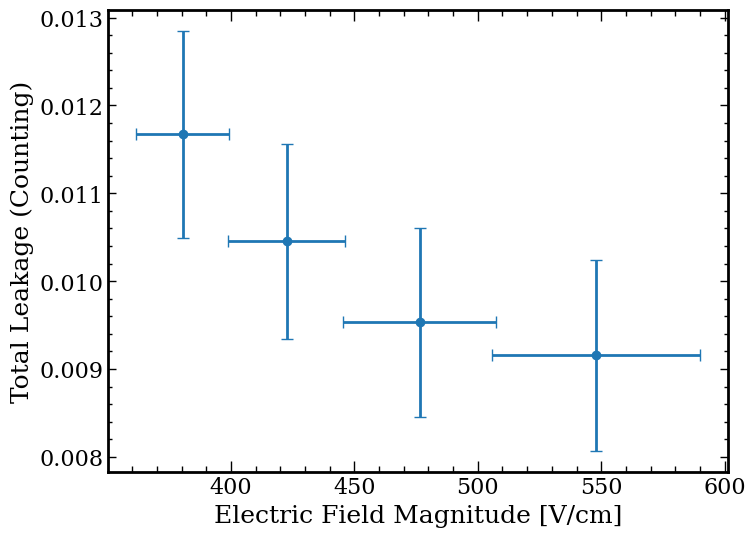}
    \includegraphics[width=0.98\columnwidth]{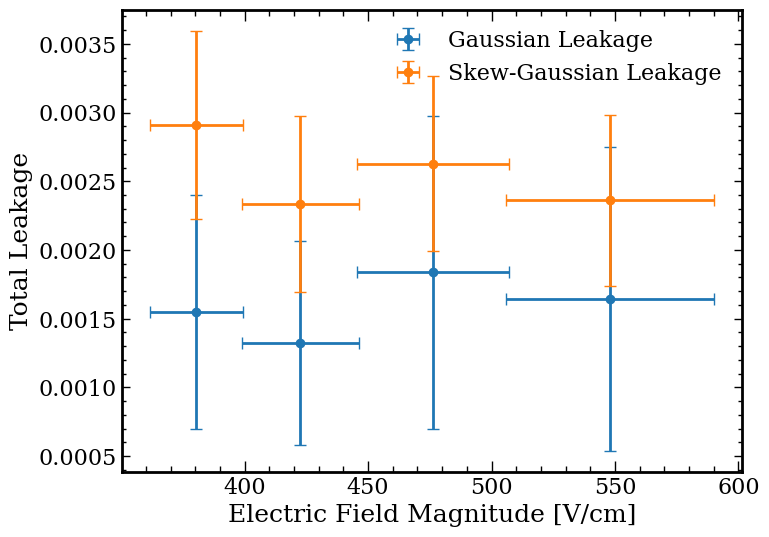}
    \caption{Top: Total leakage estimated by directly counting the proportion of events below the NR median. Bottom: Total leakage estimated by calculating the proportion of the ER events below the NR median via a Gaussian or skew-Gaussian fit.}
    \label{fig:total_leakages}
\end{figure}

\begin{figure*}
    \includegraphics[width=0.49\textwidth]{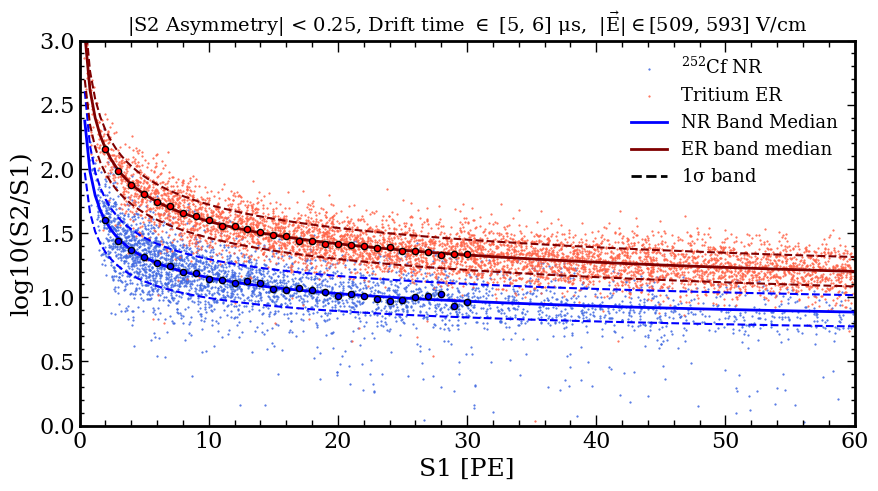}
    \includegraphics[width=0.49\textwidth]{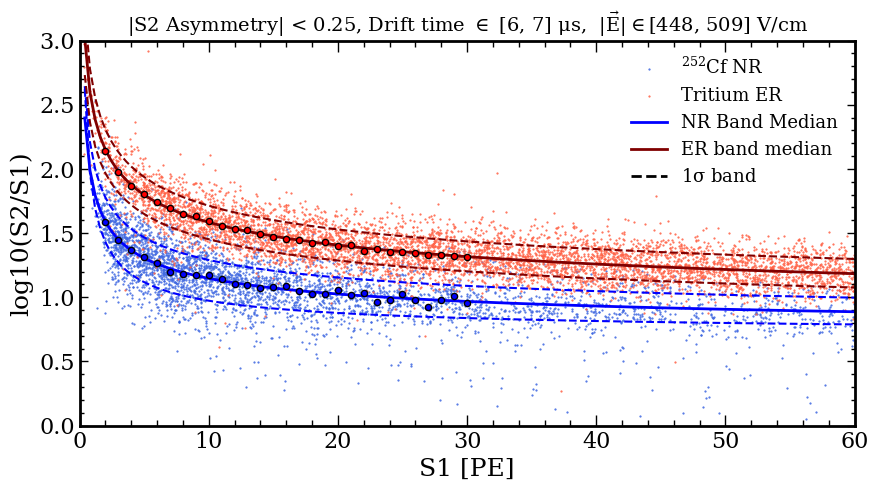}
    \includegraphics[width=0.49\textwidth]{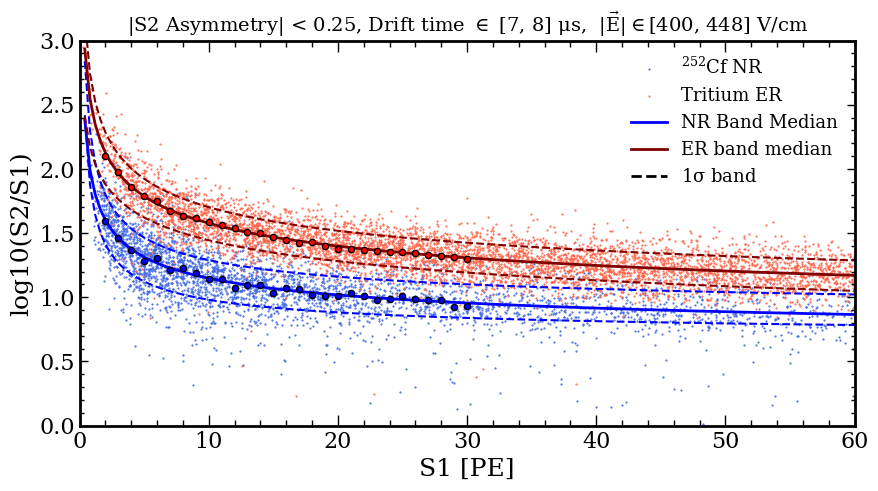}
    \includegraphics[width=0.49\textwidth]{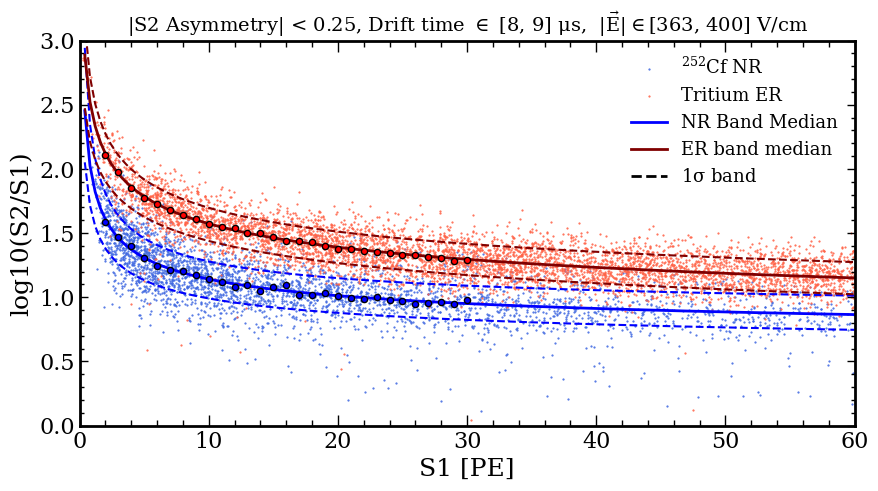}
    \caption{The ER (red) and NR (blue) scatter plots for 1~$\mu$s drift time slices from 5-6~$\mu$s. The median and $\pm1\sigma$ lines are calculated to be the 50th, 16th, and 84th percentile of the $\log_{10}(S2/S1)$ distribution for each S1 bin. In the plot, these quantile lines are fit with a power law, and are used to determine the leakages. At around 20~PE in S1, the NR band starts to show some bleed-in from ER events, likely due to the large number of gamma events from having the $^{252}$Cf source near the detector during data taking (Figure \ref{fig:er_nr_bands}).}
    \label{fig:er_nr_scatterplots}
\end{figure*}

\begin{figure*}
    \centering
    \includegraphics[width=0.45\textwidth]{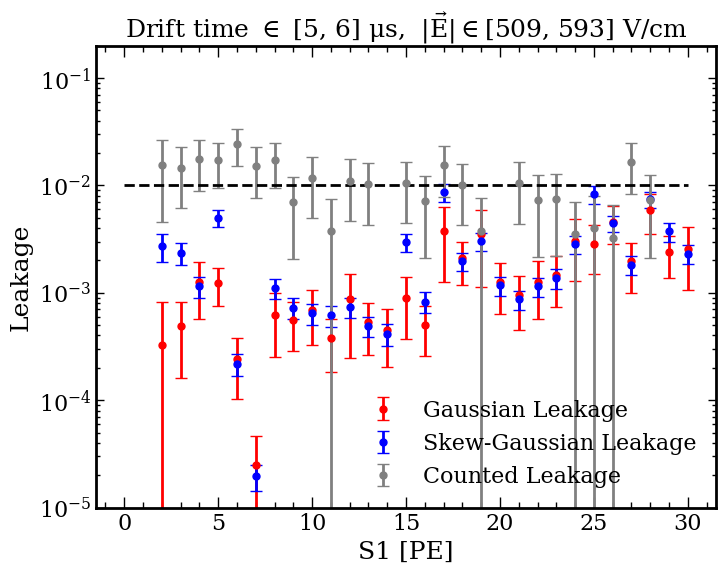}
    \includegraphics[width=0.45\textwidth]{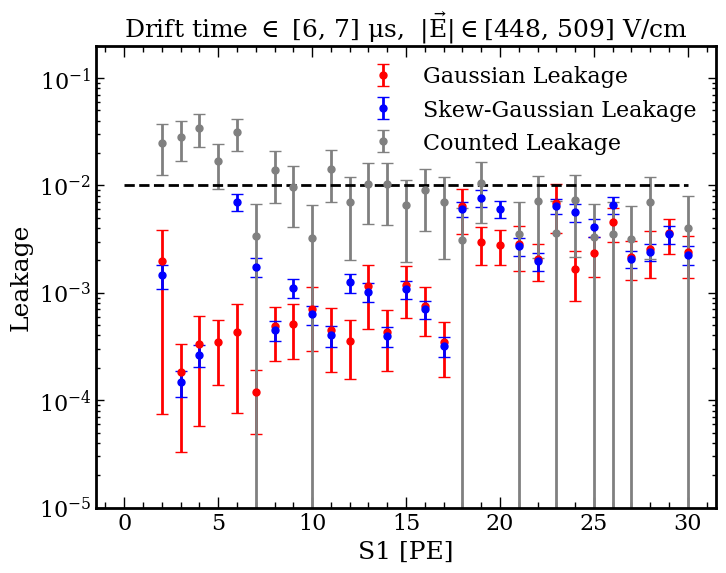}
    \includegraphics[width=0.45\textwidth]{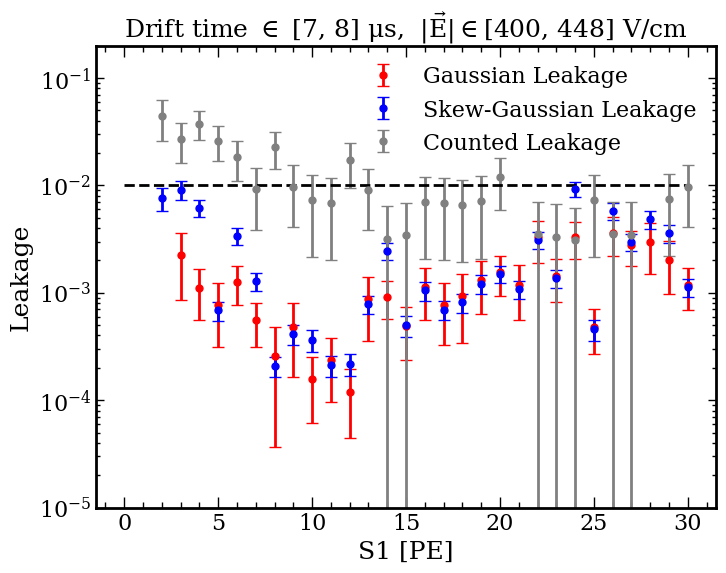}
    \includegraphics[width=0.45\textwidth]{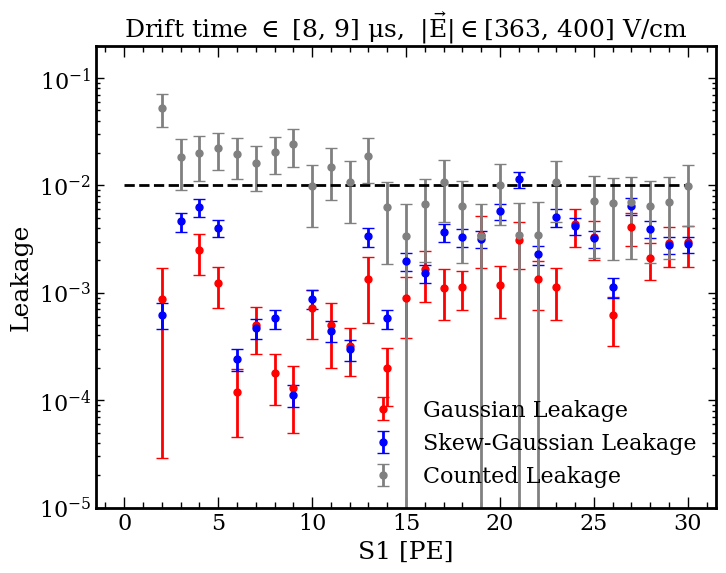}
    \caption{The leakage of ER events below the NR median for each 1~$\mu$s drift time bin as a function of the S1 area.}
    \label{fig:leakage_per_s1}
\end{figure*}

After all of the aforementioned cuts and corrections are applied, we take the tritium ER and NR events and bin them according to their drift time in 1~$\mu$s wide slices from 5~$\mu$s to 9~$\mu$s. The leakage is defined to be the proportion of tritium ER events which lie below the median of the NR band in $\log_{10}(S2/S1)$ space (Figure \ref{fig:er_nr_scatterplots}). This leakage can be calculated for all events under the NR band, or for each S1 bin. Furthermore, one may either calculate the leakage by directly counting the raw proportion of tritium events that lie below the NR median, or by fitting the $\log_{10}(S2/S1)$ distribution for the ER band in each S1 bin with either a Gaussian or skew-Gaussian then calculating the area below the NR median. When using the fitting method, the total leakage across the entire ER band up to 30 PE in S1 is the weighted average of the leakage in each S1 bin, where the weights are the number of ER events in the bin. Some example fits are shown in Figure \ref{fig:example_fits_leakage}. The leakages per S1 bin for each drift time are shown in Figure \ref{fig:total_leakages}, and the total leakage for a given drift time (i.e. electric field) slice is shown in Figure \ref{fig:leakage_per_s1}.

The total leakages calculated using the direct counting method are around 0.009-0.012, meanwhile the fitted leakages are on the order of $10^{-3}$. Although these results show that ER/NR discrimination can still be retained even with the low $g_2$ from LXe proportional scintillation, the directly counted leakages are significantly higher than those obtained in many other dual phase LXeTPCs \cite{XENON:2017sic}\cite{LUX:2020car}, which are on the order of $10^{-4}-10^{-3}$. Meanwhile, the total leakage calculated with the Gaussian or skew-Gaussian fitting is comparable to dual phase LXeTPCs. A comprehensive summary of the reported leakages from dual phase LXeTPCs is given by the NEST collaboration (Figures 13 and 14 of \cite{Szydagis:2022ikv}). While there may be other sources of leakage events, perhaps due to event reconstruction effects, we identify that one source of leakage events which leads to the high leakage obtained from direct counting is partial charge loss due to a charge insensitive volume. 

\subsubsection{\label{subsubsec:wfsim}Charge Insensitive Volume and Waveform Simulation}
\begin{figure*}
    \centering
    \includegraphics[height=0.27\textheight]{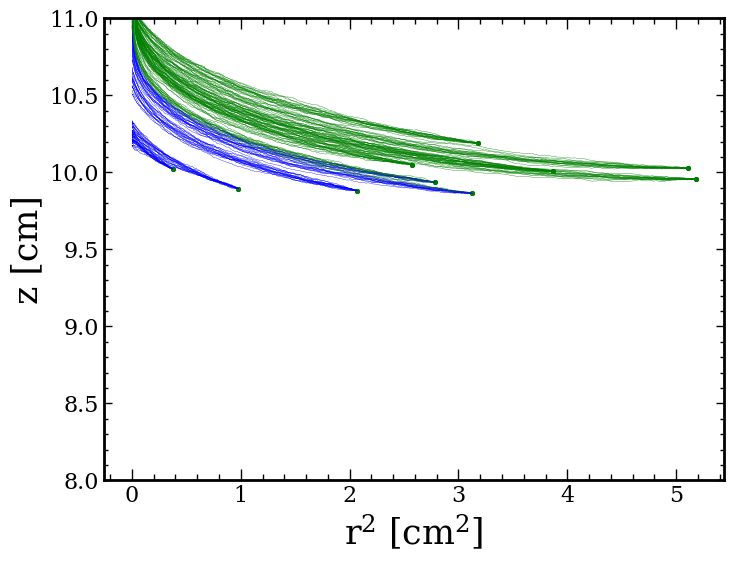}
    \includegraphics[height=0.27\textheight]{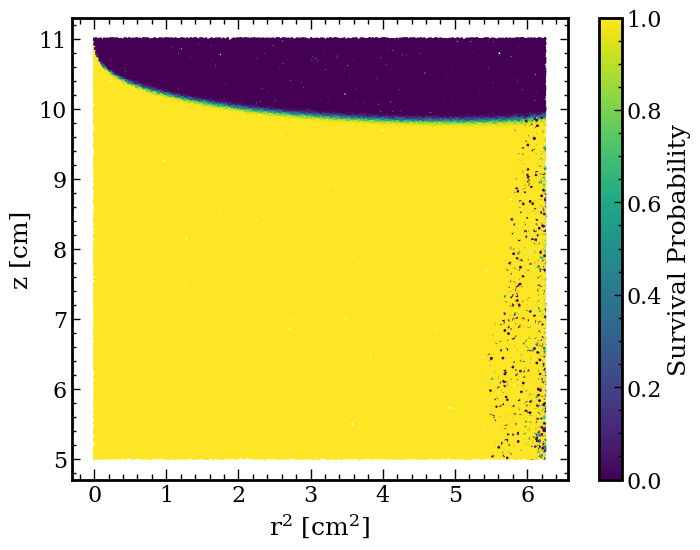}
    \caption{Left: Tracks of an electron cloud generated at a point indicated with the green dots. Green tracks show electrons which end up on the PTFE plates, blue tracks show electrons which end up on the anode. The fifth set of tracks from the bottom shows partial charge loss where some electrons strike the PTFE, and some electrons strike the anode. Right: The survival probability map projected onto $r^2$ and $z$.}
    \label{fig:partial_charge_loss_rz_tracks}
\end{figure*}

Charge-insensitive volume (CIV) induced leakage events may arise from an event generated near the CIV, in which some of the ionization electrons may diffuse onto the top and bottom PTFE plates rather than the anode. Thus these electrons are not detected, and this effect yields a smaller-than-expected S2. In order to check this hypothesis, however, we first need to simulate the full detector response. This simulation takes the energy depositions within the detector, and yields the digitized waveforms as if they were read out directly from a digitizer. Afterwards, we may transfer this simulated digitizer output through the same processing pipeline used to process real raw data. In principle, this should reproduce the same reconstruction effects that we see during real data taking. The full description of our waveform simulation is given in Appendix \ref{appendix:wfsim}. An illustration of this partial charge loss can be seen in Figure \ref{fig:partial_charge_loss_rz_tracks}.

\begin{figure}
    \centering
    \includegraphics[width=0.9\columnwidth]{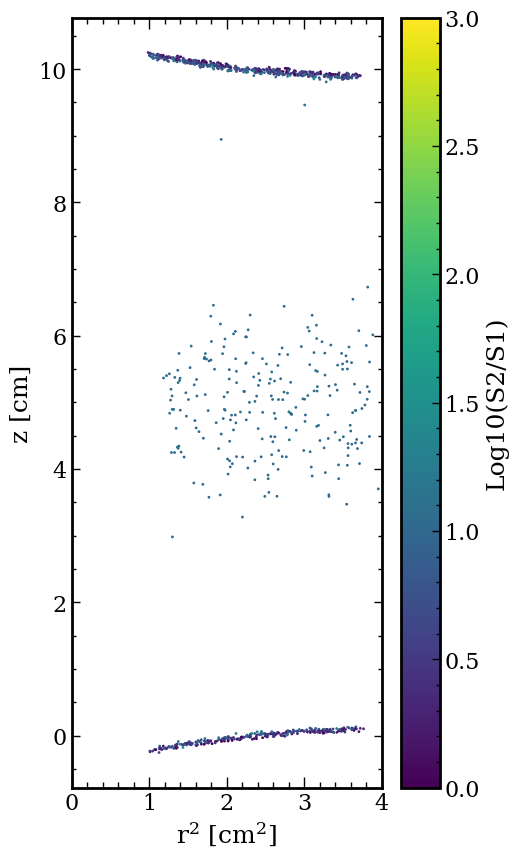}
    \caption{The monte-carlo truth positions of the simulated tritium ER events which lie below the NR median and have a drift time between 5 to 9~$\mu$s. We can see clearly that these events are largely concentrated near the border of the charge insensitive volume.}
    \label{fig:mc_truth_leakage}
\end{figure}

This waveform simulator generated the ER and NR bands shown in Figure \ref{fig:er_nr_bands}, from which we can clearly see the leakage events which lie below the majority of the ER band. Despite the fact that the partially charge insensitive volume is located near the top and bottom of the detector, an S2 asymmetry cut between -0.25 and 0.25 still does not get rid of these events. This is due to the fact that the S2 asymmetry only acts as an accurate proxy for the true $z$ position when the S2 is sufficiently large. As events near the CIV have a small S2, their S2 asymmetry does not accurately reflect their true $z$ position. This phenomenon is shown in Figure \ref{fig:mc_truth_leakage}, and is a likely component of the leakage events seen in this run. However, it is important to reiterate that this is a qualitative argument, as a true quantitative estimate for the leakage events due to the CIV would require a low uncertainty simulation across multiple inputs. These inputs include the NEST light and charge yields across a multitude of electric fields, the electric field near the top and bottom of the detector, the optical simulation, and the diffusion coefficients of electrons in LXe.

\section{\label{sec:etrains}Electron Trains in Single-phase Liquid Xenon}

There are two, hypotheses for the origin of the slow emission of electrons after a large ionization signal (i.e. the electron trains background) seen in dual-phase LXeTPCs. The first is that the electrons are trapped and slowly released (or induced to release) from impurities in the bulk of the LXe, while the second is that electrons are trapped and released near the liquid-gas interface. The LUX experiment has shown that the rates of the electron trains background decreases with the electron lifetime, indicating that these electrons are related to impurities in the bulk \cite{LUX:2020vbj}. This is supported further by the fact that both XENON1T and LUX \cite{XENON:2021qze,LUX:2020vbj} observed a higher electron trains background from events with a longer drift time (i.e. more electrons lost to impurities). However, LUX also showed that events near the top of the detector still have this background, which may suggest that some of these electrons may be still trapped near the liquid-gas interface. XENON1T also showed a weak dependence of the electron trains background due to the extraction field strength, but also did not observe a strong dependence on the electron lifetime. Furthermore, Sorensen and Kamdin showed two distinct exponential components of the electron trains: a fast component whose amplitude decreases with the extraction field, and a slow component which is related to the purity but does not decrease with the extraction field \cite{Sorensen:2017kpl}. Recently, Sorensen reported a study of PTFE flourescence, which seems to decay as a power law and may induce the release of electrons due to impurities \cite{Sorensen:2024idm}.

The LXePSC removes the liquid-gas interface entirely, which also removes the relevance of the extraction efficiency or possible trapping of electrons on impurities at the liquid-gas interface. As such, if we are able to identify single-electrons, then we can determine the role that the liquid-gas interface plays in the electron trains background. In our previous runs, the $g_2$ value was too low to distinguish single-electrons from a pileup of single photons. Now, given our $g_2$ of $\sim$3~PE/electron, we are in a position to make PMT coincidence requirements to obtain a cleaner population of single-electrons needed to calculate the electron trains rate. Furthermore, the triggerless data taking method allows for indefinite time windows, meaning we are able to look at several tens of milliseconds after an S2.

\subsection{\label{subsec:light_emission}Light Emission Limitation}

\begin{figure*}
    \centering
    \includegraphics[width=0.9\textwidth]{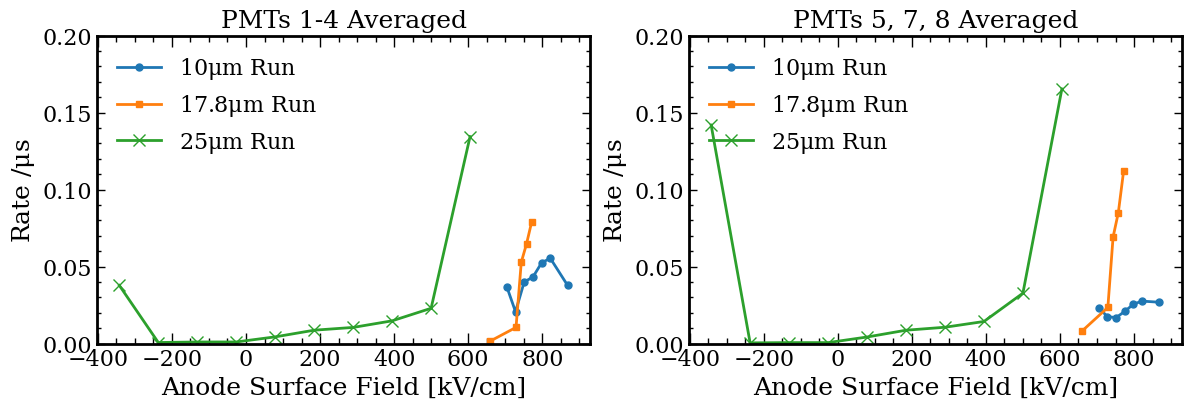}
    \caption{Light emission rates throughout the three runs of the LXePSC. These rates are calculated by counting the number of pulses in each PMT which lie 10~$\mu$s before main S1 peaks within background data. This current run is shown with the orange curve, and the operating anode voltage of 4.5~kV is shown with the second smallest anode surface field.}
    \label{fig:light_emission_rates}
\end{figure*}

Throughout our three runs, the limiting factor for our $g_2$ was the onset of spurious light emission when the anode voltage is set too high. This light emission can be seen in the example NR waveform in Figure \ref{fig:er_nr_wfs}. Once light emission rates are too high, single-photons are likely to form accidental coincidences across channels and ruin the detector's ability to reconstruct small S1s and single-electrons. From Figure \ref{fig:light_emission_rates}, we can see that thinner anode wires can reach higher electric fields before this onset happens. However, this does not necessarily mean that $g_2$ will always be larger for a thinner anode wire, as the proportional scintillation region is also smaller. Meanwhile if the wire is too thick, one needs to apply a larger voltage to achieve an electric field high enough to produce proportional scintillation in the liquid. This light emission rate also increases when the rate of energy depositions within the detector are higher, such as when a radioactive source is present as shown in \cite{Qi:2023bof}. In addition to limiting $g_2$, this light emission also acts as a background when calculating the electron trains rates.

\subsection{\label{subsec:trains_model_rates}Peak Trains Modelling and Results}

Previous experiments have shown that the rate of electrons following a large S2 follows a power law proportional to the area of the primary S2 \cite{XENON:2021qze,LUX:2020vbj}. Here, ``primary" means the S2 which generates peaks after it. However, our above ground detector has a much higher event rate than the low-background underground LXe detectors. For comparison, the event rate in LUX was 3-4 counts per second \cite{LUX:2020vbj}, which corresponds to 250-333~ms on average between each event, whereas we observed 16~ms on average between each event. Since we know that the electron trains can last $\mathcal{O}$(1~s) after an S2, the electron trains of previous S2s are likely to contribute to the current S2. In addition, we have a seemingly constant background, due in part to the spurious light emission. Therefore, for a given S2, the rate of peaks following it is modelled as: \begin{equation}
    R(\Delta t)=A_0c\Delta t^p + \sum_{i=1}^{N_{prev}} A_i c(t_i + \Delta t)^p +R_{bkg}.
    \label{eq:trains_rate_before_avg}
\end{equation} Here, $\Delta t$ is the delay time after the current S2, $c$ is a proportionality constant, $p$ is the power law power, $A_0$ is the area of the current S2, $A_i$ is the area of the $i^{th}$ previous S2, $t_i$ is the time difference between the current S2 and $i^{th}$ previous S2, and $R_{bkg}$ is a constant background rate. Here, $c$, $p$, and $R_{bkg}$ are free parameters. The rate that is actually measured is an average over many S2s. When doing this averaging, one not only needs to average over the areas of the previous S2s, but also over the time differences between the $i^{th}$ previous S2 and the current S2. Therefore the actual trains rates are fit to: \begin{subequations}
    \begin{multline}
    \langle R(\Delta t) \rangle =\langle A_0 \rangle c\Delta t^p + \\
    \langle A \rangle c\sum_{i=1}^{N_{prev}}\int_0^\infty (t_i+\Delta t)^p P(t_i)dt_i + R_{bkg}
    \label{eq:trains_model}
    \end{multline}
    \begin{equation}
        P(t_i) = \frac{t_i^{i-1}e^{-t_i/\tau}}{\tau^i(i-1)!}
    \end{equation}
\end{subequations} Here, $P(t_i)$ is the Erlang distribution, which is the distribution of the time difference between events that are $i$ events apart. $\tau$ is the average time between two consecutive events, which is approximately 16~ms. Note that $\langle A_0 \rangle \neq \langle A \rangle$, since $\langle A_0 \rangle$ is affected by the cuts applied to select for primary S2s, while $\langle A \rangle$ is a term associated with the contribution of the previous S2s' electron trains, and is thus out of the analyst's control.

\begin{figure}[!h]
    \centering
    \includegraphics[width=0.98\columnwidth]{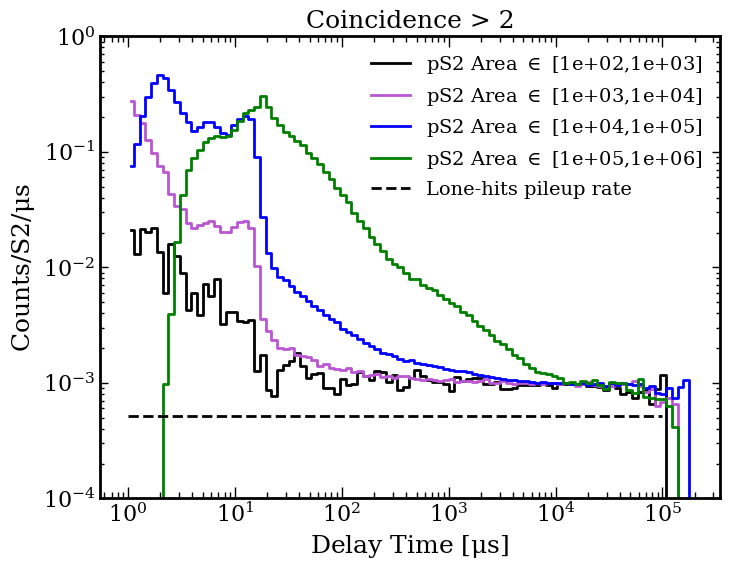}
    \caption{Trains rates for different area selections of the primary S2. All peaks used in the calculation of these rates have at least 3-fold coincidence. The green curve indicating S2s with $>10^5$~PE often contains mostly muon events which are several times wider than the typical S2. These events also distort the baseline and saturate the maximum waveform pulse length, making it difficult to see peaks immediately after it ends. The lone-hits pileup rate as indicated by the black dashed curve is obtained via the calculation described in the text.}
    \label{fig:trains_rate_diff_ps2_3_fold}
\end{figure}

\begin{figure}[!h]
    \centering
    \includegraphics[width=0.98\columnwidth]{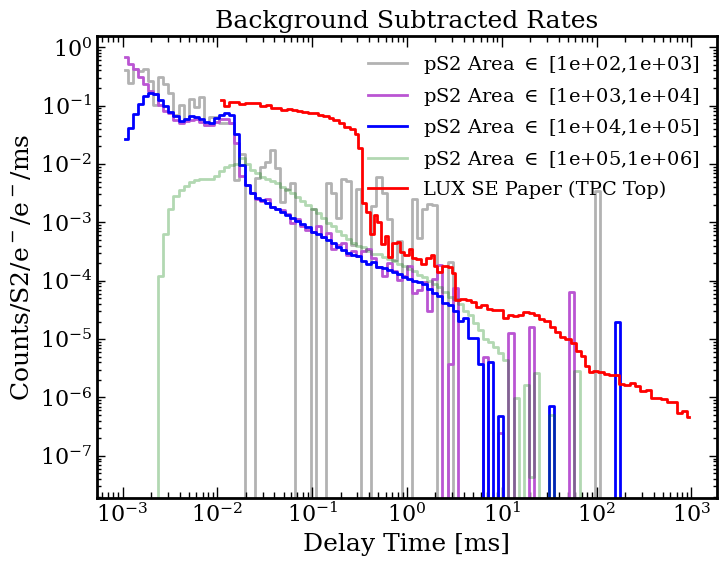}
    \caption{The trains rates with the $10^{-3}$ $\mu s^{-1}$ constant rate (Figure \ref{fig:trains_rate_diff_ps2_3_fold}) subtracted, then divided by the average primary S2 area in electrons (not PE). The statistics for peaks following a primary S2 area between 100 and 1000~PE are too low, whereas the primary S2s with an area over $10^5$~PE saturate the DAQ. Otherwise, we see that the trains rates normalized by the primary S2 area for primary S2s between $10^3$ and $10^4$ are well matched to each other. These rates are also divided by the detection efficiency for 3-fold coincident single electrons, obtained from the simulation described in the text. A more simulation-driven estimation of the electron-only part of the trains is described in the text and shown in Figure \ref{fig:trains_with_fits}, whereas the rates shown above are not reliant on simulation.}
    \label{fig:compare_trains_rate_with_lux}
\end{figure}

To calculate the train rates, we first compute the time window used to look for peaks after an S2. The start of this window is the end time of the current S2, and the end of this window is the start time of the first peak with an area over 100~PE that occurs after the current S2. We select for primary S2s which belong to single-scatter events, have an S2 area greater than 1.5 times its S1 area, and have an S1 area of at least 20~PE. This ensures that the events come from physically sensible energy depositions. The delay time of a peak within the time window is the time difference between the start of the peak and the end of its primary S2. These delay times are then binned for each S2 into a histogram, and these histograms are then added over all S2s (within a given set of cuts). For each delay time bin, we then count the amount of time windows which are greater than the given delay time bin's end, and divide each bin of the original rate histogram by this count. Doing so gives us the electron train rates as shown in Figure \ref{fig:trains_rate_diff_ps2_3_fold}.

A first look at Figure \ref{fig:trains_rate_diff_ps2_3_fold} shows some striking characteristics of the rates of small peaks after a large S2, which are similar to what is seen in the dual-phase TPCs. The most obvious is the steep drop-off around 15~$\mu$s, which is also our maximum drift time. This drop was also seen in our previous run \cite{Qi:2023bof}, and comes from the photoionization of the cathode. Next, we see that there is a component of the peak rate that \textit{increases} with the area of the primary S2, and decays on the scale of several milliseconds. The one part which differs from dual-phase TPCs is a constant background component that seems to be the same regardless of the primary S2 area. To check if the peaks within the decaying part of the trains rates are due to electrons, we can look at the area spectrum of the peaks within certain delay time bins, and compare them to the area spectrum of lone-hits which accidentally pile up to form peaks with at least 3-fold coincidence. The latter peaks are referred to as ``lone-hit pileups". To estimate the area spectrum of lone-hit pileups, we first calculate the lone-hit rate for each PMT by counting the amount of lone-hits which occur within 10~$\mu$s before the S1 of a proper S1-S2 pair. Next, for each PMT, we randomly place 161~ns wide toy pulse windows on a 1~s time window. The amount of pulses for each channel is sampled according to a Poisson distribution using the lone-hit rate for that channel, meanwhile the areas of the pulses are sampled from the channel's lone-hit area spectrum. Overlapping pulses in time are then merged, and a 3-fold coincidence requirement is placed on the resulting simulated peaks, which is used to compute the area spectrum. This same method also yields the \textit{rate} of lone-hit pileups. 

\begin{figure}[!h]
    \centering
    \includegraphics[width=0.98\columnwidth]{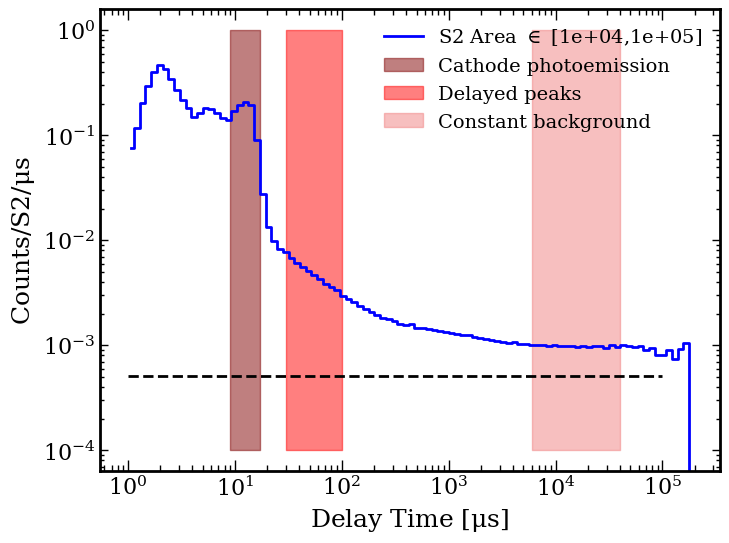}
    \includegraphics[width=0.98\columnwidth]{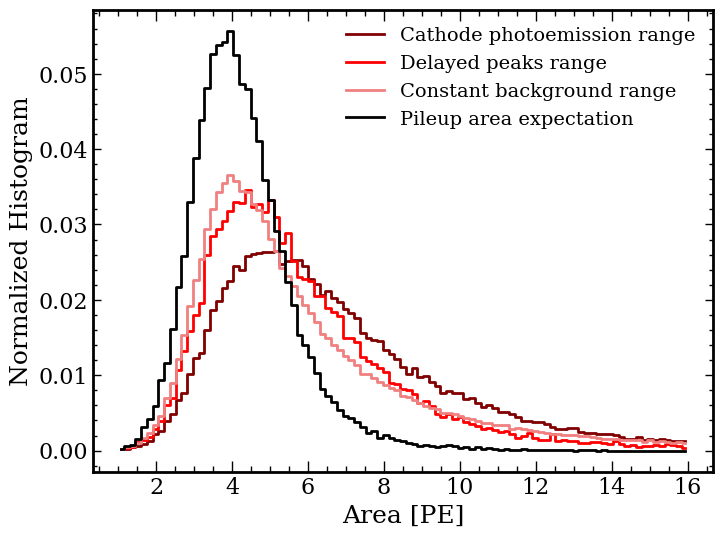}
    \caption{Top: The blue curve is the same as the blue curve in Figure \ref{fig:trains_rate_diff_ps2_3_fold} and indicates the rate of peaks after a primary S2 for primary S2s with an area between $10^4$ and $10^5$~PE. These S2s are not from excessively wide muons, and have an area which is large enough to provide a significant count of peaks over the constant background. The maroon, red, and pink shaded areas correspond to the delay time ranges used to select the peaks for which an area spectrum is computed in the bottom plot. Bottom: The area spectrum of peaks selected within the delay time regions indicated on the top plot. The pileup area expectation is the the expected area distribution of 3-fold coincidence lone-hit pileups, where the calculation is described in the text.}
    \label{fig:trains_area_spec_dt_sel}
\end{figure}

\begin{figure}
    \centering
    \includegraphics[width=0.98\columnwidth]{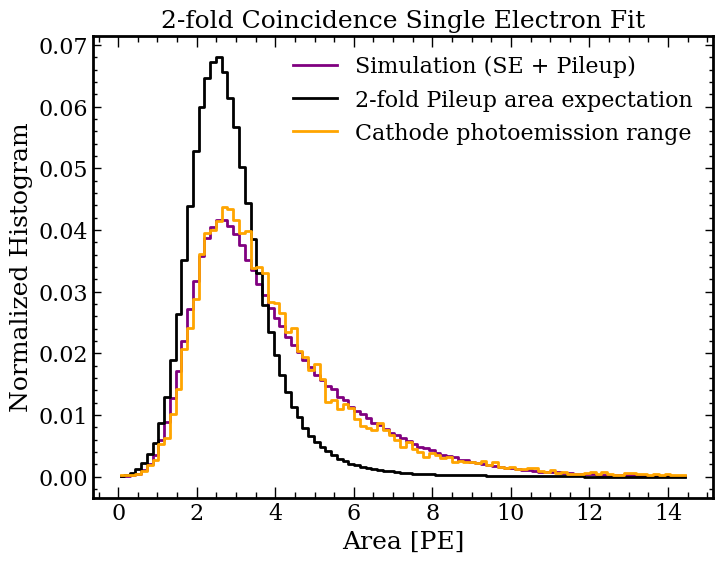}
    \caption{The two-fold coincident cathode photoemission peaks fit with a linear combination of the single-electron and two-fold lone-hit pileup area spectra. The parameters of this simulation yield an estimated $g_2=3.3$~PE/electron, which is obtained via an average of the number of photons detected per electron (excluding zero), times $1+p_{PDE}$. This figure is consistent with the 3.2~PE/electron figure in Section \ref{sec:g1g2}, which uses the old PMT gains and no S1 LCE correction.}
    \label{fig:se_fit}
\end{figure}

From Figure \ref{fig:trains_area_spec_dt_sel}, we can see that the area spectrum of the peaks in all parts of the delay time histogram extend into substantially larger areas than what is expected of lone-hit pileups. Therefore, we currently have evidence of peaks which occur after an S2, whose rate rises with its preceding S2, decays with the delay time, and whose area spectrum is characteristically different from lone-hit pileup. This set of evidence seems consistent with the notion that at least some of these peaks are due to electrons. If there are truly single and few electrons in the peaks following an S2, then we should be able to discern how many there are by fitting the area spectrum of the peaks in each delay time bin with a linear combination of the lone-hit pileup area spectrum and the $n-$electron area spectra \begin{equation}
    H(A|\Delta t) = (1-\sum_{i=1}^{N_e} r_i)p_{pileup}(A) + \sum_{i=1}^{N_e}r_i p_{i,e}(A).
    \label{eq:spec_separate}
\end{equation} Here, $H(A|\Delta t)$ and $r_i$ are the area spectrum and proportion of the peaks with a delay time around $\Delta t$, respectively. $p_{i,e}(A)$ is the area spectrum of $i$-electron peaks. To obtain $p_{i,e}$, we only need to know the parameters needed to simulate single-electrons. These parameters are $g_{nph}$, the true number of photons produced per electron before detection efficiencies, and $p_{DPE}$, the probability of double PE being emitted at the PMTs. To fit these parameters, we use the area spectrum of the two-fold coincident peaks which occur 9-17~$\mu$s after a primary S2 that has an area between $10^3$ and $10^4$~PE. The 9-17~$\mu$s delay time range selects for the cathode photoemission peaks (Figure \ref{fig:trains_area_spec_dt_sel}), in which we know there are electrons.  Furthermore, a primary S2 with an area between $10^3$ and $10^4$~PE is bright enough to provide a sufficient amount of these electrons. Meanwhile, peaks occurring near the cathode photoemission time range after primary S2s with an area larger than $10^4$~PE are skewed towards higher areas, indicating that there may be a pileup of single-electron peaks which combine to double-electron peaks. A two-fold coincidence requirement is used over the three-fold requirement to provide a higher detection efficiency towards single-electrons, albeit at the cost of a greater lone-hit pileup background. We fit the aforementioned selected peaks' area spectrum as a linear combination of the two-fold coincident lone-hit pileup area spectrum, and the simulated single-electron area spectrum which is obtained using the waveform simulation as described in Appendix \ref{appendix:wfsim}. With a photon detection efficiency of 25~\%, we estimate $g_{nph}=25.06^{+0.03}_{-0.06}$~photons/electron and $p_{DPE}=0.23^{+0.04}_{-0.02}$ (Figure \ref{fig:se_fit}).

\begin{figure*}
    \includegraphics[width=0.32\textwidth]{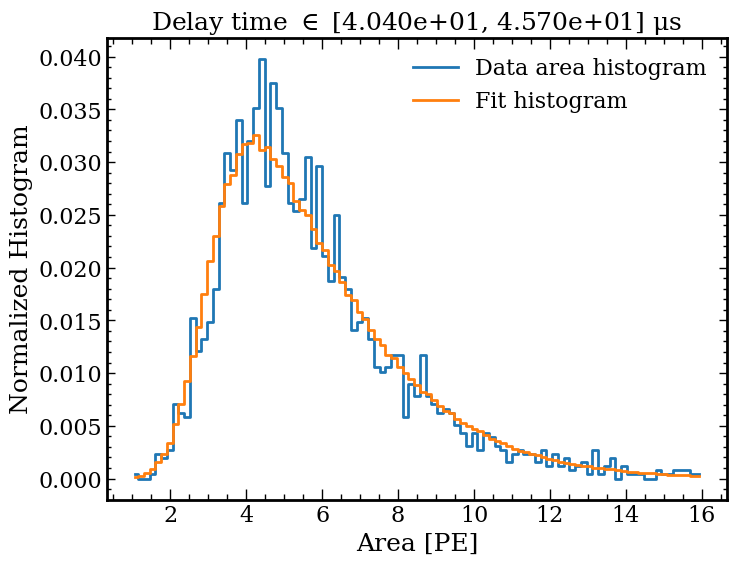}
    \includegraphics[width=0.32\textwidth]{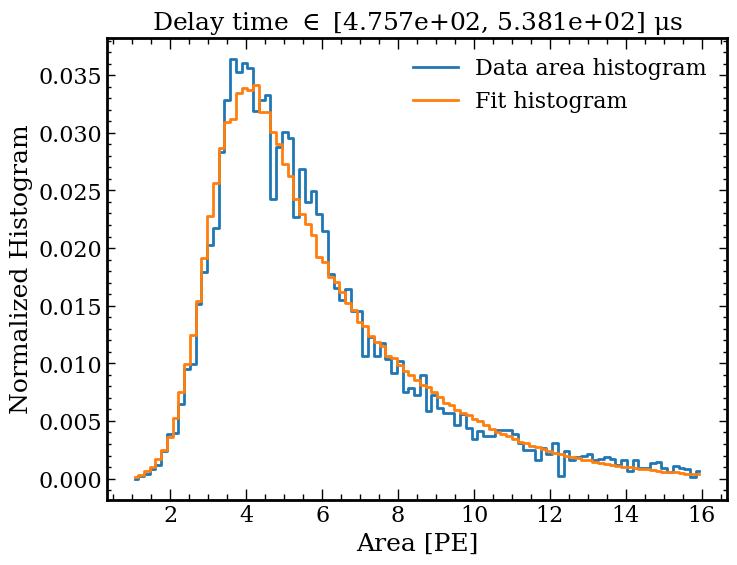}
    \includegraphics[width=0.32\textwidth]{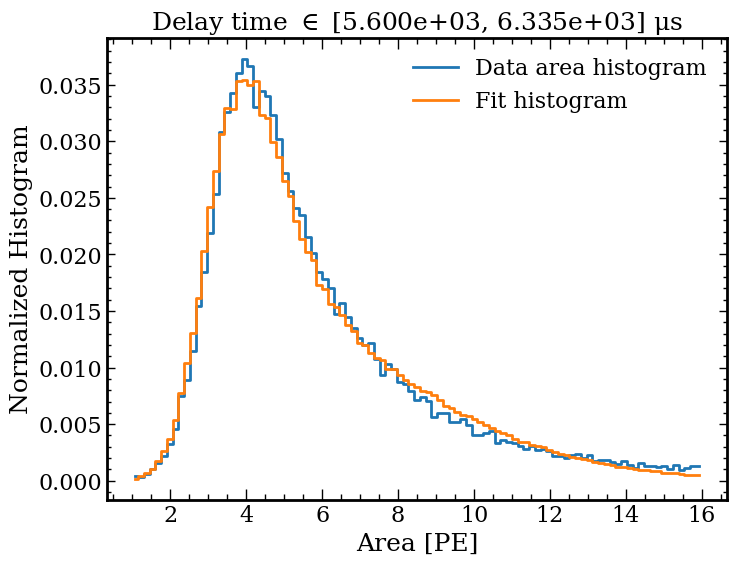}
    \caption{Example fits of the area spectrum of peaks within certain delay time bins to Eq. \ref{eq:spec_separate}. The delay time bin is indicated by the title of the plot, and the area spectrum templates are obtained via the simulation explained in the text.}
    \label{fig:example_trains_area_spec_fits}
\end{figure*}

\begin{figure}
    \centering
    \includegraphics[width=0.98\columnwidth]{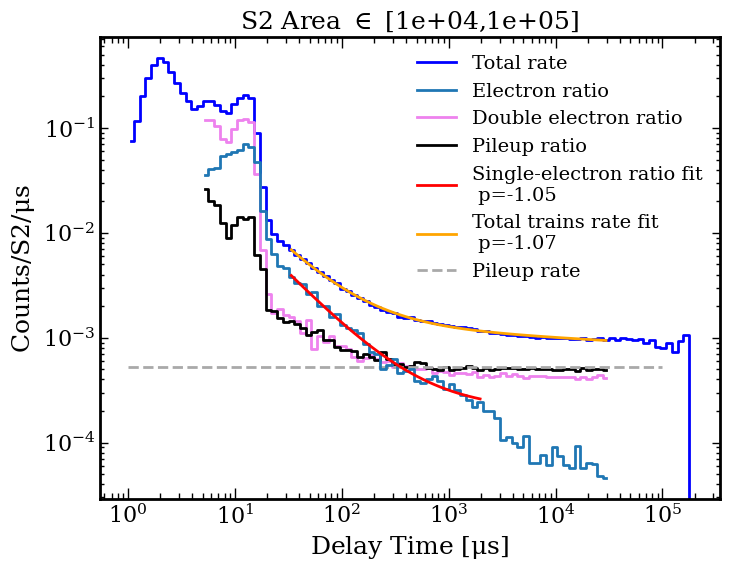}
    \caption{The peak trains split into its separate components. Whether or not we fit the single-electron portion, or the total rate with Eq. \ref{eq:trains_model}, we get a power law power of around -1, similar to the dual-phase LXeTPCs.}
    \label{fig:trains_with_fits}
\end{figure}

After obtaining template area spectra for lone-hit pileups, single-electrons, and double-electrons, we fit the area distributions of the three-fold coincident peaks following primary S2s with an area between $10^4$ and $10^5$~PE in each delay time bin to Eq. \ref{eq:spec_separate}. Example fits are shown in Figure \ref{fig:example_trains_area_spec_fits}. Multiplying the original peak trains rates by the $r_i$ and $1-\sum_i r_i$ obtained from the fit in each delay time bin gives the proportion of $i$-electron and lone-hit pileups respectively. These separated components are shown in Figure \ref{fig:trains_with_fits}. Here, we can see a clear single-electron rate that decays similarly to a power law. Furthermore, we also observe that the rate of lone-hit pileups converges to the value obtained from measuring the lone-hit rate in each PMT. However, a population of double-electrons also forms a component of the constant background, which is not understood, and whose origin is beyond the scope of this analysis. When fitting the total peak trains rate, or when fitting the estimated single-electron trains rate with Eq. \ref{eq:trains_model}, we obtain a power law power of -1.07 and -1.05 respectively. This power law power is similar to what has been observed in other dual-phase LXeTPCs \cite{Kopec:2021ccm, LUX:2020vbj, XENON:2021qze}, and indicates that at least a substantial part of the electron trains background is not due to complications at the liquid-gas interface.

To further this claim, we can integrate our trains rates over a comparable integration window to the trains rate in LUX. We choose to use the LUX results as these rates show a steep drop off after the photoionization, a power law, and an integration window with a time-range of 500~$\mu$s-3~ms -- comparable to what we have observed. A rough comparison of the trains rates to LUX is described and shown in Figure \ref{fig:compare_trains_rate_with_lux}. The integral of the electron portion of the trains between 500~$\mu$s to 3~$\mu$s (light blue line, Figure \ref{fig:trains_with_fits}), divided by the average primary S2 area in electrons as well as the 3-fold single electron detection efficiency, yields $2.4\times10^{-4}$. Meanwhile, the same calculation done on the LUX train rates (within the top 5~cm of the detector) yields $5.9\times10^{-4}$. Both of these rates are on the same order of magnitude, despite the fact that the electron lifetimes and maximum drift times are not the same.

\section{\label{sec:conclusion}Conclusion and Discussion}

The LXePSC has demonstrated that ER/NR discrimination for low-energy recoils is possible using proportional scintillation to amplify charge directly in liquid, despite the fact that the single-electron gain ($g_2$) is around 10 times smaller than that of dual-phase detectors. This is due to the fact that even with a $g_2$ of a few PE/electron, $g_2$ is still over ten times larger than $g_1$. As such, the LXe proportional scintillation process may be useful for future dark matter searches in order to achieve a perfect extraction efficiency and to mitigate sagging effects of large electrodes. One such example of this technology uses a microstrip-coated quartz plate as an electrode \cite{Martinez-Lema:2023zjk}, which retains the typical vertical drift region with a constant field that is usually seen in TPCs. Furthermore, the light-emission issue would need to be resolved first before such a detector is deployed in a dark matter search. With regards to low-mass dark matter or CE$\nu$NS searches using the ionization-only channel, our results indicate that eliminating the liquid-gas interface alone is not likely to reduce the electron trains background that limits the sensitivity of such searches. Nevertheless, given the aforementioned recent results of P. Sorensen \cite{Sorensen:2024idm}, a cylindrical geometry with photosensors that surround the detector inner barrel may provide sufficient photocoverage to efficiently detect scintillation light without the use of PTFE reflectors, and thus may still mitigate the background of single electrons.

\section{Acknowledgements}
This research is sponsored by the US Defense Advanced Research Projects Agency (DARPA) under grant number HR00112010009, the content of the information does not necessarily reflect the position or the policy of the Government, and no official endorsement should be inferred. Jianyang Qi is supported by the High Energy Physics Consortium for Advanced Training (HEPCAT) graduate fellowship from the Department of Energy grant DE-SC0022313.

\newpage
\nocite{*}
\bibliography{apssamp}

\appendix

\section{\label{appendix:satcor}Saturation Correction}
\begin{figure}
    \centering
    \includegraphics[width = 0.99\columnwidth]{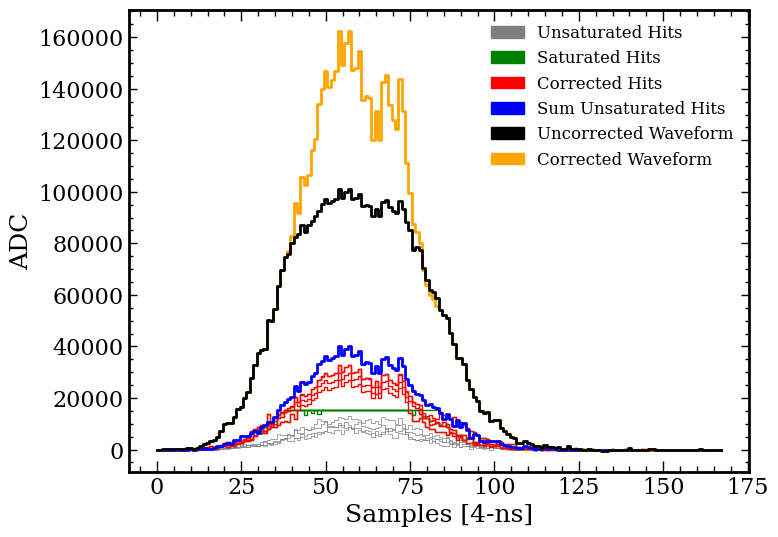}
    \caption{A demonstration of the saturation correction. The blue waveform is the sum of the unsaturated channels shown in gray. This is then scaled to replace the saturated parts of the saturated channel waveforms as shown in green.}
    \label{fig:satcor}
\end{figure}

We call the sum of the channel waveforms for the unsaturated channels, $wf_{us}$. Afterwards, we replace the saturated portion of the saturated channel waveforms, $wf_s$, with $wf_{us}$ scaled to match the height of the first saturated sample of $wf_s$. An example of this can be seen in Figure \ref{fig:satcor}. This saturation correction does not work when all channels are saturated. For this reason, we chose to estimate $g_1$ and $g_2$ by selecting for events with a drift time between 5~$\mu$s and 9~$\mu$s, which corresponds to an $r$ between approximately 1.1-1.8~cm. Below 5~$\mu$s, a significant portion of the S2s have all channels saturated, and above 9~$\mu$s (1.8~cm), we expect there to be field inhomogeneities that are difficult to model (see Fig. \ref{fig:e_vs_r}). The drift time to $r$ conversion is done using NEST \cite{szydagis_m_2022_7061832} drift velocities and the COMSOL field simulation.

\section{\label{appendix:er_nr_corrections}Corrections to the ER and NR Events}

\begin{figure}
    \centering
    \includegraphics[width=0.9\columnwidth]{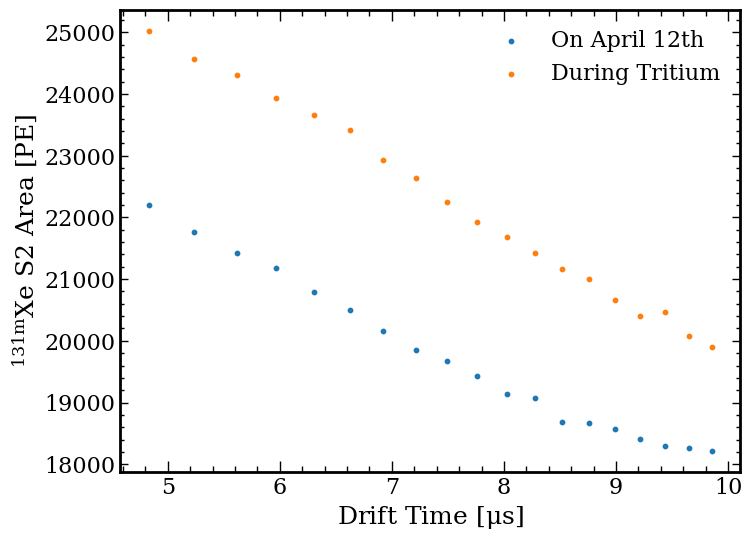}
    \caption{The difference in the central S2 area of the $^{131\rm m}$Xe line for the data taken on April 12th versus the data taken during the tritium calibration. The large upward shift in the S2 area during tritium data taking is due to the increase in circulation speed.}
    \label{fig:trit_data_cor}
\end{figure}

Before taking tritium data, we placed the $^{252}$Cf source near our detector for two days which gave us activated xenon lines in-situ with the tritium data. We can thus directly compare the S2 area of the $^{131\rm m}$Xe line for each drift time bin during tritium data taking, to that of the data taken on April 12th, 2023 (see Figure \ref{fig:trit_data_cor}). As such, the corresponding correction factor is \begin{equation}
    C_{S2} = \frac{S2_{^{131m}Xe}(t_d; \text{April 12th})}{S2_{^{131m}Xe}(t_d; \text{Tritium data taking})}.
\end{equation}

During NR data taking, we did not have activated xenon lines yet, so we use the 39.6~keV gamma line from inelastic neutron scattering off $^{129}$Xe \cite{TIMAR2014143} to compare the area of S2s during the NR data taking to that of April 12th. The 39.6~keV peak during NR data taking is selected by estimating the CES distribution using $g_1=0.158$~PE/photon and $g_2=3.35$~PE/electron. These values are from the activated xenon data using the old set of PMT gains without the S1 LCE correction. Note the difference from the quoted $g_1$ and $g_2$ values in Section \ref{sec:g1g2}, as those values are obtained using the S1 LCE correction as a systematic uncertainty. Selecting the events with a CES between 29 and 46 keV (Figure \ref{fig:40kev_line_selection}) gives us the desired 39.6~keV peak.

\begin{figure}
    \centering
    \includegraphics[width=0.9\columnwidth]{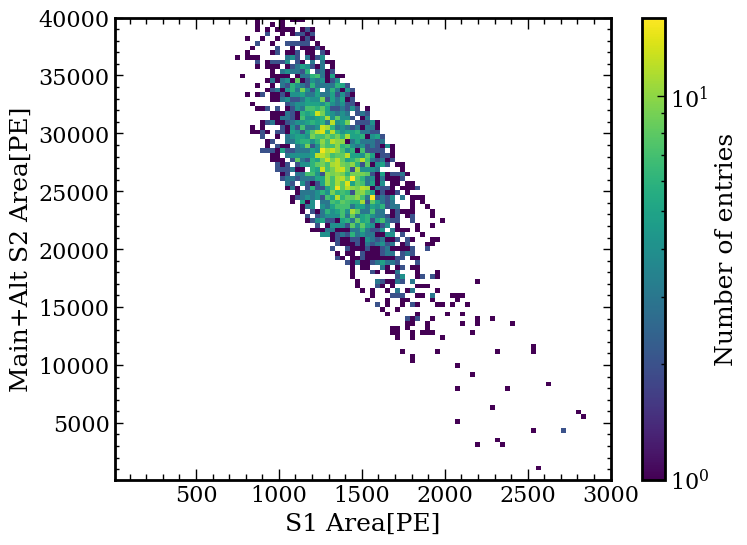}
    \includegraphics[width=0.9\columnwidth]{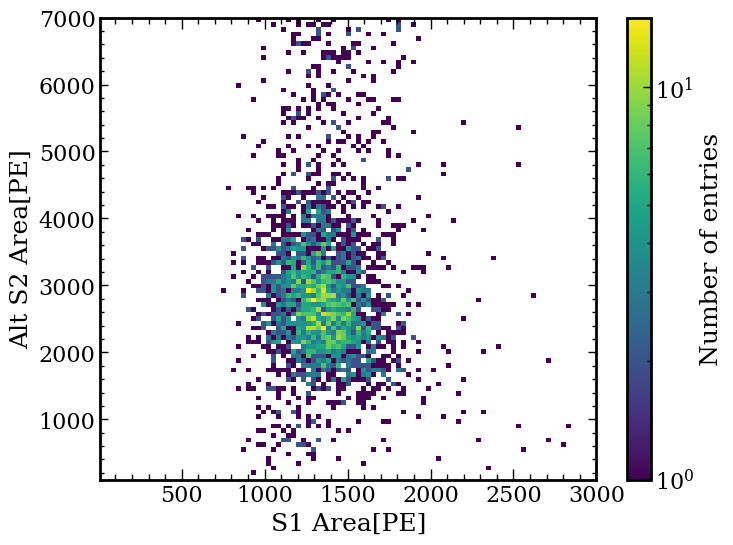}
    \caption{Top: The total (S2, S1) distribution of the 196~keV and 39.6~keV gammas together whose S1s are merged together but S2s are separate. Bottom: The alternate S2 vs main S1 distribution of the events from the above plot, these S2s should correspond to the 39.6~keV line.}
    \label{fig:xe_act_40_kev}
\end{figure}

At first glance, there does not seem to be a 39.6~keV peak in the activated xenon data taken on April 12th. However, we can exploit the fact that the decay of $^{129\rm m}$Xe into its ground state is actually a two-step process, with the first step releasing a gamma line at 196~keV which decays with an 8.8~day half life, and the second step which decays with a half life of 1~ns and releases our desired 39.6~keV line \cite{TIMAR2014143}. The mean free path of the 196~keV peak in LXe is approximately 8.6~mm while that of the 39.6~keV peak is $\sim$142~$\mu$m \cite{136421}. As there is only $\sim$1~ns between the first and second step, the S1s of these gammas will be merged. However, if these two gamma rays deposit their energy in different radial locations that are far enough from each other, their S2s will be separated. Nonetheless, the total S1 and \textit{sum} of alternate and main S2 should still combine to give the same 236~keV (Figure \ref{fig:xe_act_40_kev}). Therefore, if we select for events with exactly two S2s and one S1, compute the CES with the main and alternate S2 summed together, and select for the 236~keV peak, the alternate S2 of this event will correspond to the S2 of the 39.6~keV gamma line.

\begin{figure}
    \centering
    \includegraphics[width=0.9\columnwidth]{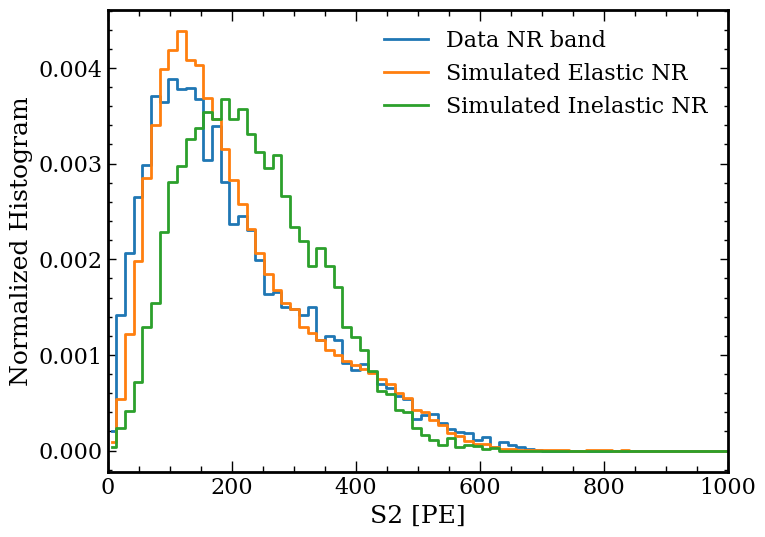}
    \caption{Comparison of the S2 distributions between data NRs to elastic and inelastic simulated NRs. The data consists mostly of elastic NRs and seems to match well with the elastic NR simulation, which justifies the use of the inelastic NR S2 distribution to perform the deconvolution with the apparent 39.6~keV line.}
    \label{fig:nr_inl_sim}
\end{figure}

\begin{figure}
    \centering
    \includegraphics[width=0.9\columnwidth]{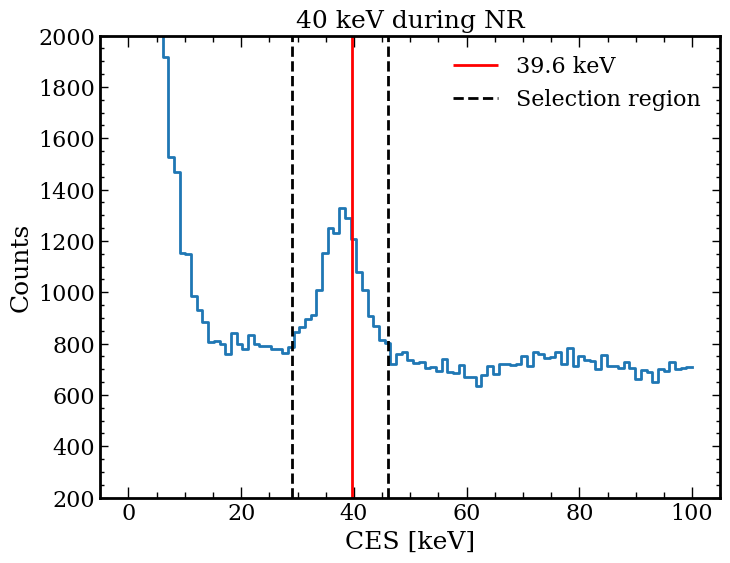}
    \includegraphics[width=0.9\columnwidth]{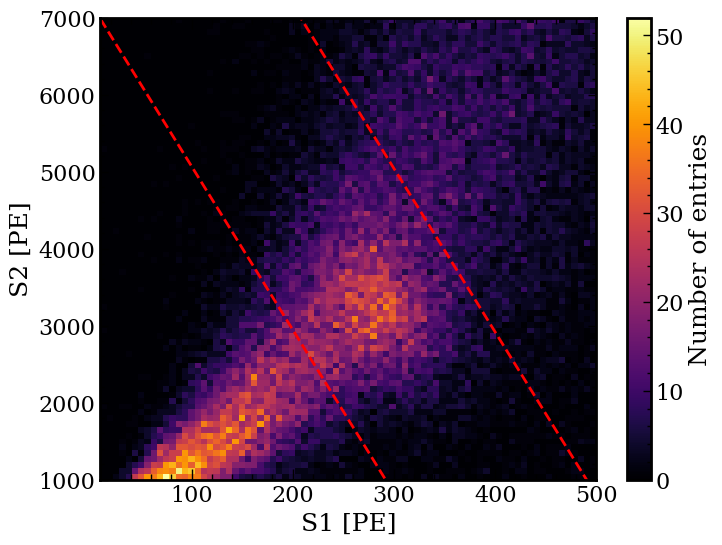}
    \caption{Top: The estimated CES of the NR data with an assumed $g_1$ and $g_2$ given in the text and an electron lifetime of 80~$\mu$s. Bottom: The 39.6~keV line apparent in (S2, S1) space. The red lines are the indicated energy selection of 29-46~keV.}
    \label{fig:40kev_line_selection}
\end{figure}

\begin{figure}
    \centering
    \includegraphics[width=0.9\columnwidth]{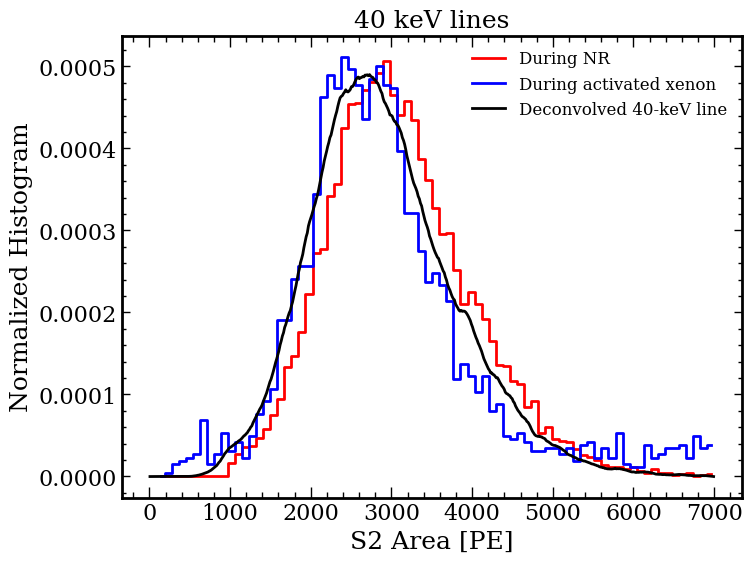}
    \caption{The S2 distributions for the 39.6~keV line during NR data taking and April 12th. We see that the peaks of the distributions are not notably different. The deconvolution was performed using Wiener deconvolution to handle the noise of the measured distributions.}
    \label{fig:deconvo_40kev}
\end{figure}

Before comparing the S2 distributions of the 39.6~keV lines taken during NR data to those taken on April 12th, one needs to recognize that the aforementioned, apparent, 39.6~keV line during NR data is actually merged with \textit{inelastic} NRs. To account for this, we model the S2 distribution of the apparent 39.6~keV line as \begin{multline}
    f(S2_{\gamma+NR}) = \int p(t_d)P(S2_\gamma = S2_{\gamma+NR} - S2_{NR}|t_d)\\\times P(S2_{NR}) dt_d dS2_{NR}
\end{multline} In other words, the observed S2 distribution of the merged 39.6~keV line and inelastic NR, for each drift time, is equal to a convolution of the true 39.6~keV S2 and the inelastic NR S2 distributions. The \textit{elastic} NR S2s are easily identified using the cuts described in the following paragraph, but we do not have such a clean population of inelastic NRs. As such, we rely on the simulation described in Appendix \ref{appendix:wfsim} to infer the inelastic NR S2 distribution, using the elastic NR S2 distribution as a validation to the simulation. The comparison of the NR simulation to data is shown in Figure \ref{fig:nr_inl_sim}. To estimate the true $P(S2_\gamma)$, we deconvolve $f(S2_{\gamma+NR})$ with $P(S2_{NR}|t_d)$ for each 1~$\mu$s wide drift time bin fron 5~$\mu$s to 9~$\mu$s. While there are backgrounds of non-39.6~keV gamma ray events, these seem to be relatively flat as shown in Figure \ref{fig:40kev_line_selection}, and thus the peak of the S2 distribution should still indicate the peak of the merged 39.6~keV line and NR. We see from Figure \ref{fig:deconvo_40kev} that the estimated 39.6~keV lines from both the NR data and April 12th data are peaked at approximately the same spot. As such, we do not apply a correction to the S2 for the NR data.  Furthermore, we do not expect that the S1 changes appreciably with time.

Both tritium ER events and NR events are low energy events on the order of a few keV$_{ee}$. As such, their primary background is due to accidental coincidences (ACs) of lone S1s and lone S2s. To cut against these, we first only keep events with an S2 pulse-width which rises with the drift time due to electron diffusion as this selects for physically sensible events. Secondly, we cut away events that occur within a ``noisy" region of data taking. This includes events within 20~$\mu$s of a previous event, and events with more than 3 other S1s within a 10~$\mu$s window centered around the primary S1. Lastly, we require that the S2s have light that is seen by all eight PMTs, as this seems to drastically reduce the AC background. In addition to cutting away AC events, we also use the same asymmetry cut as described in Section \ref{sec:g1g2} and require that the alternate S2 area is less than 1\% of the main S2 area. Lastly, we define a parameter called the S2 undershoot, which is the ratio of the negative area of an S2 waveform, to the positive area of the S2 waveform. S2s with an undershoot ratio of greater than 0.03 are cut, which leaves 98\% of tritium events and 94\% of NR events.

\section{\label{appendix:wfsim}Waveform Simulation}

\begin{figure}
    \centering
    \includegraphics[width=0.9\columnwidth]{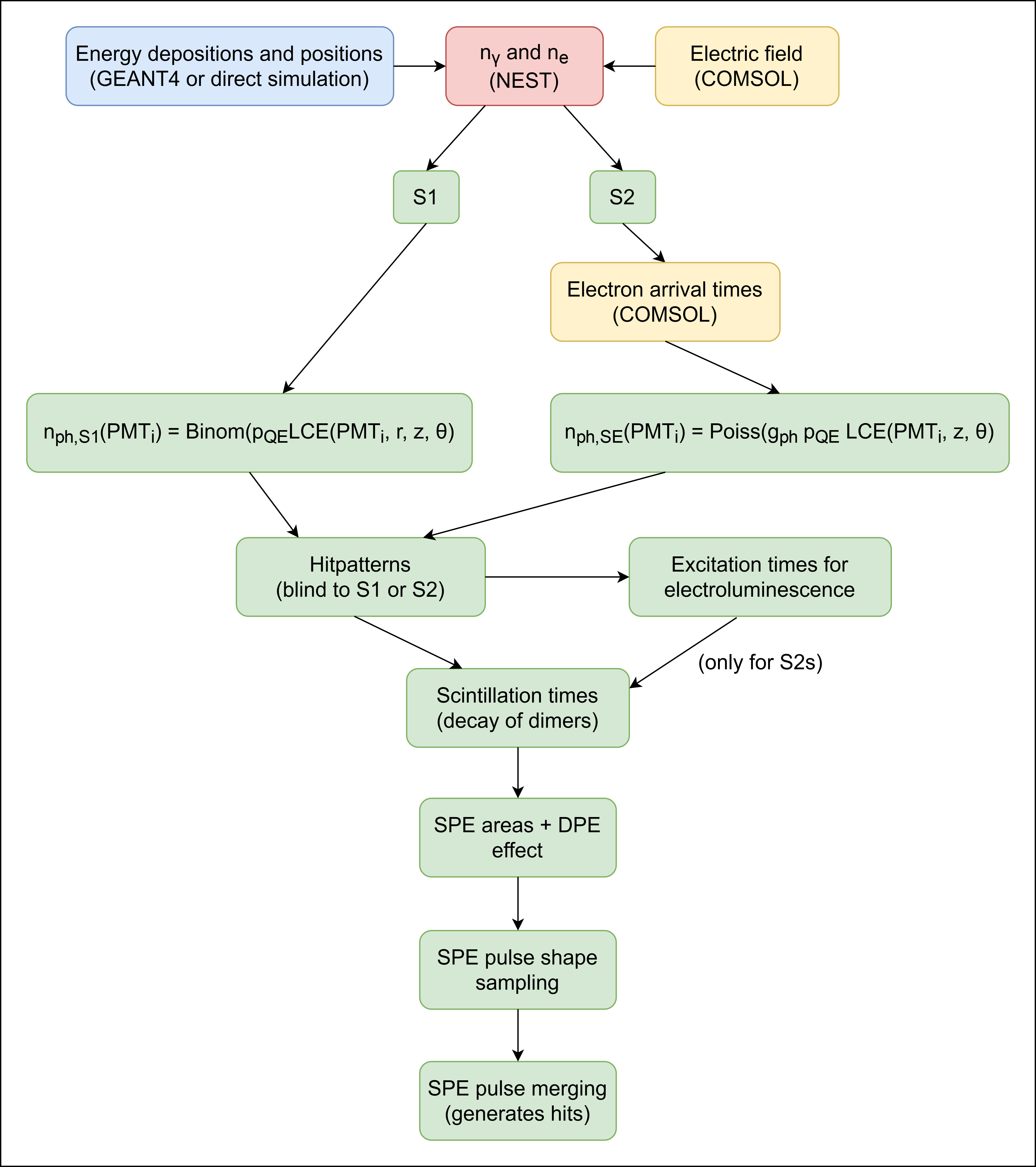}
    \caption{Steps in the full-chain waveform simulation.}
    \label{fig:wfsim_steps}
\end{figure}

For external sources such as fast neutrons from $^{252}$Cf, we use GEANT4 to simulate the positions and energy depositions within our detector. Meanwhile, the energies of uniformly distributed sources -- such as tritiated methane -- can be sampled directly according to their energy spectrum, while their positions of those energy depositions can be sampled uniformly within the detector volume. The locations of an event are then used to obtain the local electric field at the interaction site. Afterwards, the local electric field, energy, and interaction type (ER or NR) are fed into NEST \cite{szydagis_2018_1314669} to obtain the number of photons, $n_\gamma$, and number of electrons, $n_e$, from each energy deposition.

For S1s, the number of photons seen by each PMT is sampled from a binomial distribution with a probability $p_{QE}LCE(PMT_i, r, z, \theta)$ where $p_{QE}$ is the total PMT efficiency, and $PMT_i$ is the $i^{th}$ PMT.\begin{equation}
    n_{ph, S1} \sim Binom(p = p_{QE}LCE(PMT_i, r, z, \theta), N = n_\gamma)
\end{equation}

For S2s, before we obtain the number of photons and the times at which those photons are produced, we first need to drift and diffuse the electrons from the initial interaction site to the anode. For each time step, $dt$, we sample the position step of each electron according to the following model:
\begin{subequations}
\begin{equation}
    \Delta r_L \sim Gauss(\mu = v_d(|\textbf{E}|)dt, \sigma = \sqrt{2D_L(|\textbf{E}|)dt})
\end{equation}
\begin{equation}
    \Delta r_T \sim Gauss(\mu = 0, \sigma = \sqrt{2D_T(|\textbf{E}|)dt})
\end{equation}
\end{subequations} Here, $\Delta r_L$ and $\Delta r_T$ are the position steps that are longitudinal and transverse to the local electric field at the position of the electron, respectively. $v_d$ is the drift velocity obtained from NEST, $D_L$ and $D_T$ are the longitudinal and transverse diffusion coefficients from Njoya et. al. \cite{Njoya:2019ldm} and EXO-200 \cite{EXO-200:2016qyl} respectively. The transverse unit vector is uniformly sampled within a plane perpendicular to the local electric field. This electron transport simulation then gives a 3-D map of: the mean drift time, spread in drift time, distribution of final electron z-positions along the anode, and the electron survival probability for an electron cloud. In this instance, the electron survival probability is the probability that the electron drifts to the anode, and nowhere else. The charge loss due to the attachment to impurities is also accounted for by assuming an electron lifetime, and sampling the number of electrons from a binomial distribution with $p=e^{-t_d/\tau_e}$. An example of partial charge loss can be seen in Figure \ref{fig:partial_charge_loss_rz_tracks}.

Each electron undergoes electroluminescence, and the photon gain, $g_{ph}$ is the number of photons produced per electron that reaches the anode. The number of photons seen by each PMT follows a Poisson distribution of \begin{equation}
    n_{ph,SE}(PMT_i) = Poiss(g_{ph}p_{QE}LCE_{S2}(PMT_i, z, \theta)
\end{equation} Note here that $LCE_{S2}$ is a 2-D map around the anode, as the S2 light is produced only microns away from the surface of the anode when estimated using Eq. \ref{eq:r_prop_scint} with 412~kV/cm as the threshold electric field from Aprile et. al. \cite{Aprile:2014ila}.

After the hitpatterns are obtained for both the S1 and S2 pulse, we sample the number of triplets and singlets using the singlet to triplet ratio of 0.042 as measured by LUX \cite{LUX:2018zdm}. From here, the singlet and triplet de-excitation time is sampled from an exponential distribution according to their respective lifetimes. Each photon is then converted into one or two PE depending on the double PE probability. Finally, the areas and shapes of the PEs are sampled according to the single PE width and single PE waveform templates obtained from the PMT calibration.

\end{document}